\documentclass[10pt,aps,twocolumn,superscriptaddress,nofootinbib,tightenlines,floatfix,longbibliography]{revtex4-2}
\usepackage{amsmath,amssymb}
\usepackage{graphicx}
\usepackage{color}
\usepackage{siunitx}
\usepackage{hyperref}
\hypersetup{
    colorlinks=true,       
    linkcolor=blue,          
    citecolor=blue,        
    filecolor=blue,      
    urlcolor=blue           
}

\usepackage{soul}

\def\abs#1{ \left| #1 \right| }
\def\fluc#1{ \left\langle #1 \right\rangle }
\def\eqref#1{{(\ref{#1})}}
\def\Tr{\textrm{Tr}}

\def\mn{{\mu\nu}}

\begin{document}

\title{Comparisons and Predictions for Collisions of deformed $^{238}$U nuclei at $\sqrt{s_{NN}} = \qty{193}{\giga\electronvolt}$}

\author{Nicolas Miro Fortier}
\email[]{nicolas.fortier@mail.mcgill.ca}
\affiliation{Department of Physics, 
	McGill University, 3600 University Street, Montreal, QC, H3A 2T8, Canada}

\author{Sangyong Jeon}
\email[]{sangyong.jeon@mcgill.ca}
\affiliation{Department of Physics, 
	McGill University, 3600 University Street, Montreal, QC, H3A 2T8, Canada}

\author{Charles Gale}
\email[]{charles.gale@mcgill.ca}
\affiliation{Department of Physics, 
	McGill University, 3600 University Street, Montreal, QC, H3A 2T8, Canada}

\date{\today}

\begin{abstract}
	We present comparisons to experimental data along with predictions of observables for U+U and Au+Au
    collisions at 193 and 200 GeV respectively, using a multistage theoretical and computational framework consisting
     of boost-invariant IP-Glasma initial state, MUSIC hydrodynamics, and a hadronic transport 
     cascade generated by iS3D \& SMASH. Two different Woods-Saxon parametrizations were used for both systems \cite{osti_6477756,Ryssens_2023}, allowing for comparisons within our model.
     Our results show great agreement with existing anisotropic 
     flow measurements from RHIC~\cite{Adamczyk_2015,Adam_2019}. We provide predictions for 
     differential flow observables as well as multiparticle correlations and transverse-momentum-flow 
     correlations. When possible, we compare our predictions to results from Au+Au collisions at 200 GeV 
     to properly outline the effects of deformation in the initial state on final state observables. 
\end{abstract}
\maketitle

\section{Introduction\label{sec:intro}} 

Heavy-ion collisions at the Relativistic Heavy Ion Collider (RHIC) and the Large Hadron
Collider (LHC) have provided remarkable insights into the fundamental properties of matter
under extreme conditions~\cite{Busza_2018}. One of the most intriguing phenomena observed
in these collisions is the creation of quark gluon plasma (QGP), a state of matter
characterized by the deconfinement of quarks and gluons~\cite{Pasechnik_2017}. The QGP,
produced in the aftermath of collisions, exhibits collective behaviour reminiscent of a
nearly perfect fluid. This remarkable hydrodynamic behaviour allows for the conversion of
initial state anisotropies into final state observables, providing a unique window into
not only the dynamics of the QGP, but also nuclear structure and its limits.

In studies involving spherically symmetric nuclei, researchers have extensively
investigated the effects of specific subsets of initial state anisotropies resulting from
nuclear geometry fluctuations. However, a significant gap remains in our understanding of
the impact of initial state anisotropies originating from deformed nuclei, such as
$^{238}$U. These deformed nuclei introduce a wider range of anisotropies, presenting an
opportunity to explore unique combinations of fluctuations and test certain
phenomenological hypotheses within the QGP~\cite{PhysRevC.61.034905}.

Notably, our study predicts that central collisions involving prolate-shaped nuclei (such
as $^{238}$U) should lead to a negative correlation between $v_2$ and $p_T$. In contrast,
this correlation is observed to be positive at all centralities in collisions of spherical
nuclei~\cite{Schenke_transverse,Aad_2019}. Indeed, the prolate geometry of $^{238}$U
causes higher eccentricity ($\varepsilon_2$) events to generate lower $\fluc{p_T}$ and
vice-versa, inducing an anti-correlation of the two observables. These collisions also
generate higher energy densities in ultra-central configurations compared to spherical
nuclei. This extra energy density is expected to have observable effects on various
crucial observables, including elliptic flow and jet quenching~\cite{Heinz_2005}, which
serve as key probes to characterize the properties of the QGP, as well as collective
properties and momentum correlations
\cite{giacalone2021impact,jia2022shape,bally2022evidence}.

While considerable efforts have been made in studying heavy-ion collisions involving
spherical nuclei~\cite{Adam_2016,Acharya_2020,Adare_2011,Adams_2005}, experimental data
for collisions involving deformed nuclei (such as $^{238}$U) remains limited. Furthermore, new studies 
have found that nuclei which had historically been modeled spherically (such as $^{197}$Au) may actually be
deformed~\cite{Ryssens_2023,giacalone2021impact}.
Therefore, not only do collisions of deformed
nuclei present an exciting opportunity for testing our understanding 
of QGP dynamics, but `deformity' may be more common than we believed in the nuclear world. 
To test these hypotheses, it is crucial to develop a comprehensive framework that can incorporate
all types of fluctuations at all stages.

In this study, we employ an up-to-date, comprehensive and well-motivated theoretical
framework to both compare to available data from STAR~\cite{Adamczyk_2015} and extract
predictions for a wide range of observables in U+U and Au+Au collisions at $\sqrt{s_{NN}} =
\qty{193}{\giga\electronvolt}$ \& $\qty{200}{\giga\electronvolt}$ respectively. Both
collision systems will be studied using two distinct Woods-Saxon parametrizations each.
Simulations begin with the IP-Glasma model, which is based
on the Colour Glass Condensate (CGC) framework~\cite{GELIS_CGC,Iancu_2004} and provides
realistic event-by-event colour fluctuations as well as pre-equilibrium flow. IP-Glasma
has been successful at reproducing key observables across both the energy and collision
system spectra
\cite{Gale:2012rq,Gale:2013da,Schenke_2020,heffernan2023bayesian,McDonald_2017,heffernan2023earlytimes,mcdonald202331d}.
MUSIC, a relativistic viscous hydrodynamic simulation which incorporates bulk and shear
viscosities~\cite{MUSIC_Schenke, Ryu_2015}, then takes over the evolution of the system.
Once the density and temperatures of the QGP drop sufficiently, the system is
`frozen-out', meaning that fluid cells which were previously governed by hydrodynamics are
converted into hadrons before being propagated. This final stage of the simulations is
taken up by iS3D~\cite{mcnelis2020particlization} \& SMASH~\cite{Weil_2016}, two models
which combine to generate hadronic cascades. The inner workings of the various phases of
our simulations will be discussed in detail in section \ref{sec:theo}.

This paper follows the following structure: section \ref{sec:theo} examines the different
stages of our physical model in varying detail, ranging from an in-depth discussion of
IP-Glasma in subsection \ref{subsec:ipg} to a more cursory look at hadronic transport
model SMASH \ref{subsec:is3dsmash}. Section \ref{sec:obs} details the different studied
observables and their working definitions, as well as theoretical expectations for their
sensitivity to the initial state when possible. These include centrality selection
(\ref{subsec:cent}), flow analysis methods (\ref{subsec:flow}) and
transverse-momentum-flow correlations (\ref{subsec:rho}). We then show our model's
comparative and predictive capabilities and results in section \ref{sec:res}. The paper
will end with a summary and conclusion, presented in section \ref{sec:conc}.

\section{Theoretical Model\label{sec:theo}}
\subsection{IP-Glasma\label{subsec:ipg}}

Historically, the first hydrodynamic models of heavy-ion collisions were developed 
with an emphasis on the fluid dynamics of QGP
\cite{Venugopalan:1990jt,Sollfrank:1996hd,Kolb:1999it,Morita:1999vj,Nonaka:2000ek,Kolb:2000fha,
Kolb:2000sd,Huovinen:2001cy,Hirano:2001yi,Morita:2002av,Hirano:2002ds,Huovinen:2006jp}.
The initial states used by these first models 
were mainly geometric in nature~\cite{Miller_2007}. Once the relevance and success of early heavy-ion collision
simulations were established, the need for a more physically accurate and detailed initial
state model became apparent. IP-Glasma, a QCD- and saturation-based model, was first
introduced in 2012~\cite{Bjorn_gluon} and quickly became the standard in the field. It is
based on the Colour Glass Condensate (CGC) effective field theory
\cite{McLerran_1994,PhysRevD.49.3352,Iancu_2002} and classical gluon production
\cite{Gyulassy:1997vt,Kovchegov:1997ke,Krasnitz_1999,Krasnitz_2003,Dumitru:2001ux}.

\begin{figure}
    \includegraphics[width=\linewidth]{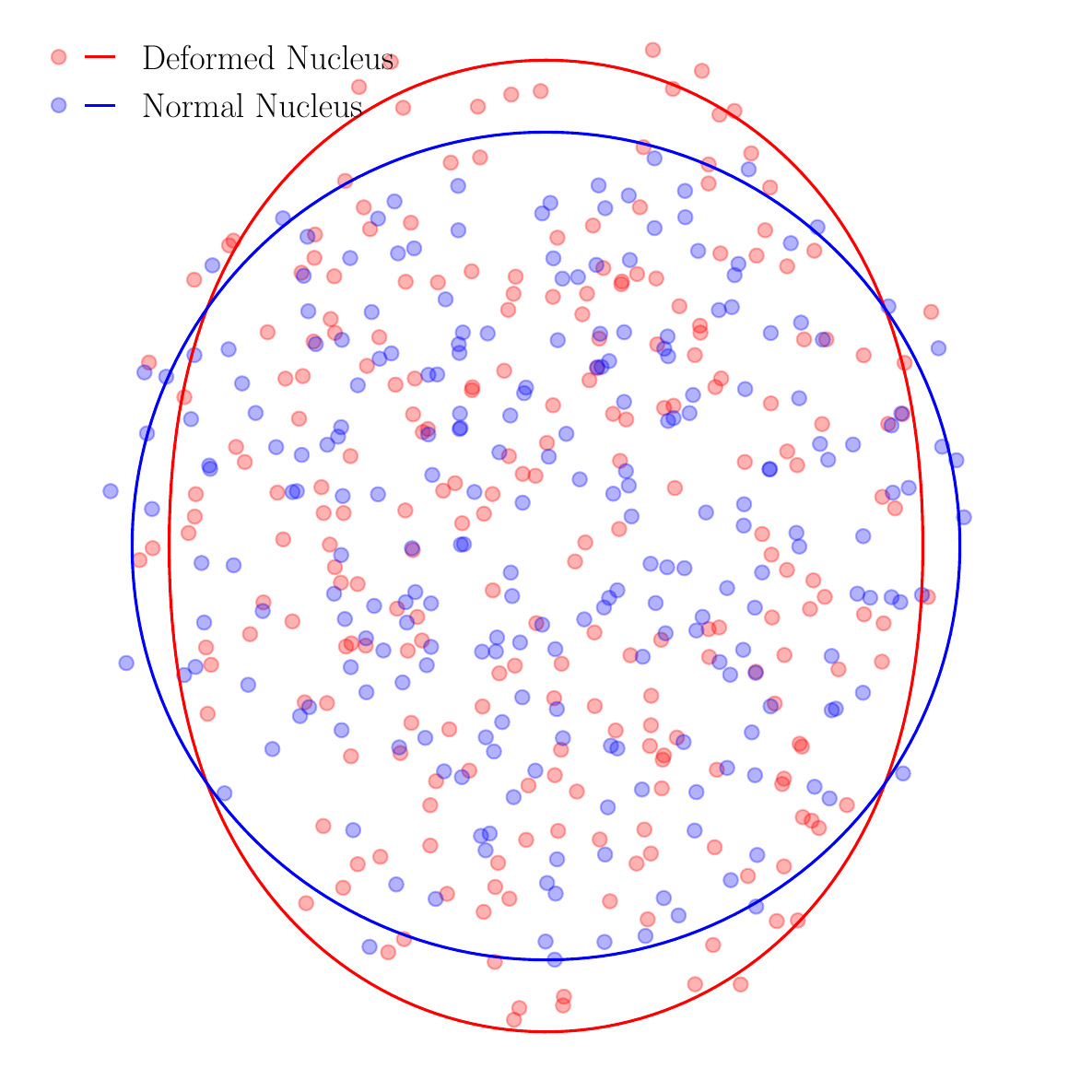}
    \caption{\label{fig:compara} Comparing a regular and deformed Woods-Saxon distribution 
    with same base radius $R_0$ and skin depth $a$ (set to that of Prev $^{238}$U in Tab.~\ref{tab:params}). The deformed distribution generates and oblong 
    (or pill-shaped) profile. The rotational symmetry axis is the long vertical axis.}
\end{figure}

To model heavy-ion collisions, one must first generate the nuclei. In this study, 
we use the deformed Woods-Saxon distribution, given by
\begin{gather}
    \rho(r,\theta) = \frac{\rho_0}{1 + \exp \left(\frac{r - R(\theta,\phi)}{a}\right)} \\
    R(\theta, \phi) = R_0\left( 1 + \sum_{l = 2}^{l_{max}} \sum_{m=-l}^{l} \beta_l^m Y_l^m(\theta,\phi) \right) \label{eq:ws}
\end{gather}
to generate nucleon configurations. Here, $\rho_0$ denotes the nuclear density (which is calculated such that the Woods-Saxon distribution integrates to 1), $R_0$ is
the unmodified nuclear radius and $a$ is the nuclear skin depth. Real parameters $\beta_l^m$ 
multiply the spherical harmonic functions $Y_l^m(\theta,\phi)$ and generate
deformation about the $x$ and $y$ axes. The parameters used in this study of $^{238}$U and 
$^{197}$Au are presented in Tab.~\ref{tab:params}. We see that all deformed parametrizations 
include quadrupole ($l=2$) and hexadecapole ($l=4$) moments, while others associated with 
triaxiality $\gamma$ are only included for the Deformed $^{197}$Au parametrization, as 
no evidence currently suggests that such parameters are relevant for $^{238}$U.
A comparison between an undeformed and a deformed nucleus using the parameters
used in this study is presented in Fig.~\ref{fig:compara}.

\begin{table}
\caption{\label{tab:params} Deformed Woods-Saxon parameters used for sampling nuclei 
according to Eq.~\ref{eq:ws}, taken from \cite{osti_6477756} (Prev $^{238}$U and Spherical Au) 
and \cite{Ryssens_2023} (New $^{238}$U and Deformed Au).
}
\begin{ruledtabular}
    \begin{tabular}{ccccc}
     &New $^{238}$U & Prev $^{238}$U & Spher $^{197}$Au &
     Def $^{197}$Au\\
    \hline
    $R_0$ (fm) & 7.068 & 6.874 & 6.37 & 6.62 \\
    a (fm) & 0.538 & 0.556 & 0.535 & 0.519 \\
    $\beta_2^0$ & 0.247 & 0.2802 & 0 & 0.098 \\
    $\beta_2^2$ & 0 & 0 & 0 & 0.076 \\
    $\beta_4^0$ & 0.081 & -0.0035 & 0 & -0.025 \\
    $\beta_4^2$ & 0 & 0 & 0 & -0.018 \\
    $\beta_4^4$ & 0 & 0 & 0 & -0.018 \\
    \end{tabular}
    \end{ruledtabular}
\end{table}

The Woods-Saxon distribution is a simplistic yet effective model for sampling nucleon
positions, as evidenced by its widespread use in the field. However, as with any simple
model, one must understand its limitations. The $\beta_l^m$ deformation parameters are
model-dependent, in so far as their extraction must figure in the Woods-Saxon
distribution. Also, the distribution itself does not account for important nucleon-nucleon
correlations, which more recent studies have shown to have sizable effects
in generating physically
accurate nuclei configurations~\cite{ALVIOLI2009225,dalal2022nucleonnucleon}. Therefore,
it should be understood that all of the results presented in this paper carry an inherent
systematic uncertainty.

Following nucleon sampling, the impact parameter for the event is sampled from
\begin{gather}
    P(b)db = \frac{2b}{b^2_{\mathrm{max}} - b^2_{\mathrm{min}}}db
\end{gather}
where $b_{\mathrm{max}} = \qty{8}{\femto\meter}$ and $b_{\mathrm{min}} = \qty{0}{\femto\meter}$. 
The boundaries were fixed with the goal of rejecting as few events as possible due to their peripherality 
since deformation effects are only tangible in central collisions of deformed nuclei. The nuclei 
configurations are then shifted symmetrically by ${b}/{2}$ in the $x$-direction. The spatial distribution 
of nucleons is then projected into the transverse plane.

At this point, IP-SAT~\cite{Kowalski_2003}, the impact parameter dependent dipole saturation model, takes over. 
Its contribution will be in providing the saturation scale $Q_s$ at all points in the transverse plane, 
which then allows us to sample colour charges and initialize the system's colour gauge fields. To do so, 
IP-SAT first models the nuclear thickness function as
\begin{gather}
    T(\boldsymbol{x}) = \frac{e^{-\boldsymbol{x}^2 / 2B_G}}{2 \pi B_G} \label{eq:single} \\
    T_{A}(\boldsymbol{x}) =\sum_{i = 1}^{A} T(\boldsymbol{x} - \boldsymbol{x_i})\label{eq:thick}
\end{gather}
where $A$ represents the current nucleus' number of nucleons and 
$B_G = \qty{4.0}{\giga\electronvolt^{-2}}$ is extracted from a fit to DIS data 
\cite{PhysRevD.87.034002}. With the thickness function for each nucleus in hand, 
we solve
\begin{gather}
    \frac{2\pi^2}{N_c}T_{A,B}(\boldsymbol{x})r_s^2 xg(x,\mu^2(r_s^2))\alpha_s(\mu^2(r_s^2)) = 1 \label{eq:satscale}
\end{gather}
for $Q_s^2 = {2}/{r_s^2}$, where $N_c = 3$ is the number of colours permitted and
$T_{A,B}(\boldsymbol{x})$ is the thickness function described in Eq.\eqref{eq:thick} 
for the projectile (\textit{A}) and target (\textit{B}) nuclei. $xg(x, \mu^2)$, the 
density of gluons at a given scale $\mu$ and momentum fraction $x$, is initialized as
\begin{gather}
    xg(x, \mu_0^2) = A_g x^{\lambda_g} (1 - x)^{5.6} \label{eq:dens}
\end{gather}
with $A_g = 2.308$, $\lambda_g = 0.058$ and $\mu_0^2 = \qty{1.51}{\giga\electronvolt^2}$.
Eq.\eqref{eq:dens} is then evolved to all other values of $\mu^2$ using the leading-order
DGLAP equation~\cite{Altarelli:1977zs,Dokshitzer:1977sg,GelisCode}. The scale $\mu$ itself
is related to the saturation dipole size (and scale) by
\begin{gather}
    \mu^2 = \frac{4}{r_s^2} + \mu_0^2 = 2Q_s^2 + \mu_0^2
\end{gather}
which sets the scale in the leading-order QCD running coupling constant given by
\begin{gather}
    \alpha_s(\mu^2) = \frac{12\pi}{(33-2N_f)\ln\left( \tfrac{\mu^2}{\Lambda_{QCD}}\right)}.
\end{gather}
where $N_f$ is the number of quark flavours, set to $4$ in our simulation. 
Eq.\eqref{eq:satscale} is itself extracted from the Glauber-Mueller dipole cross-section 
\cite{Mueller:1989st}. Solving for $Q_s$ must be done iteratively given the intricate interdependence 
of the various functions ($xg$, $\alpha_s$) and variables ($x$, $r_s$, $\mu$, $Q_s$).

Once $Q_s^2$ is determined, the colour charge distribution for the projectile nucleus, for instance, 
can be sampled from the following colour correlator
\begin{gather}
    \fluc{\rho^a_A(\boldsymbol{x})\rho^b_A(\boldsymbol{y})} = 
    g^2\mu^2_A(x,\boldsymbol{x})\delta^{ab}\delta^2(\boldsymbol{x} - \boldsymbol{y}),
\end{gather}
where $x$ is the momentum fraction currently considered and $Cg^2\mu_A = Q_s$\footnote{It
is important to note here that $\mu_A \neq \mu$. $\mu$ is the intrinsic energy scale at
hand, while $\mu_A$ is the scale of colour charge fluctuations, which is related to the
saturation scale.}. We are therefore sampling from a Gaussian of width proportional to the
saturation scale. The proportionality constant $C$ is determined phenomenologically and
was set to $0.505$ in this study.

This sampled colour charge distribution is of great importance, as it acts as a source for the 
small-\textit{x} gluon fields which comprise the CGC. The CGC action,
\begin{gather}
    S_{CGC} = \int d^4x \left(-\frac{1}{4}F^a_{\mn}F^{\mn}_a + J^{\mu a}A^a_{\mu} \right)
\end{gather}
contains a current term $J^{\mu a}$ which sources the colour gauge fields $A_{\mu}^a$. The 
convention for the covariant derivative and, therefore, the field strength tensor used throughout 
this paper is
\begin{gather*}
    D_\mu = \partial_{\mu} + igA_{\mu} \\ 
    F_{\mn} = \partial_{\mu}A_{\nu} - \partial_{\nu}A_{\mu} + ig\left[A_{\mu},A_{\nu}\right].
\end{gather*}
The corresponding Classical Yang-Mills (CYM) equation is
\begin{gather}
    \left[D_{\mu}, F^{\mn}\right] = J^{\nu} 
\end{gather}
where
\begin{gather}
    J^{\nu} = \rho_A(\boldsymbol{x})\delta^{\nu +}\delta(x^{-})
\end{gather}
Here, the two $\delta$-functions indicate a right-moving source on the light cone. Note the
move to light-cone coordinates $x^\pm = (t\pm z)/\sqrt{2}$,
meaning that our sources (large-\textit{x} partons)
are travelling at the speed of light. In the pre-collision phase, the CYM equations in the
$A_- = 0$ gauge are
\begin{gather}
    \nabla^2_{\perp}A^a_+ = -\rho^a \label{eq:poiss}
\end{gather}
which are Poisson equations for each colour index. The more physical light-cone gauge gluon fields 
can be obtained by a gauge transformation
\begin{gather}
    A^A_{\mu} = V^A A_\mu V^{A\dagger} - \frac{i}{g}V^A\partial_{\mu}V^{A \dagger} \label{eq:wils}
\end{gather}
where
\begin{gather}
    V(x^-,\boldsymbol{x}) = \mathcal{P}\exp\left( -ig\int_{-\infty}^{x^-} dy^- A_-(y^-,\boldsymbol{x}) \right)
\end{gather}
In this 2D setting, only the transverse components $A^A_x$ and $A^A_y$ are non-zero.

Once the pre-collision colour gauge fields have been determined, the collision takes place; both 
fields are combined such that the forward light cone has the following initial fields
\begin{gather}
    A_{i} = A^A_{i} + A^B_{i} \label{eq:init}\\
    E^{\eta} = ig\left[ A_i^A, A_i^B \right],
\end{gather}
where $i = x, y$.
The dynamics inside the forward light cone are best described using $\tau - \eta$
coordinates, which we will use in the rest of this work. To obtain the other components of
the chromo-electric field $E^i$, we must solve the covariant form of Gauss' law,
\begin{gather}
    \left[D_i, E^i \right] + \left[ D_{\eta}, E^{\eta} \right] = 0. \label{eq:gauss}
\end{gather}
Since derivatives in $\eta$ vanish in our boost-invariant implementation, we are left with 
$\left[D_i, E^i \right] = 0$, which has trivial solution $E^i = 0$: the initial transverse 
chromo-electric fields are therefore set to 0~\cite{kovner1995gluon,PhysRevD.52.6231}.

Once the initial post-collision fields are settled, we evolve the whole system using the 
sourceless CYM equations,
\begin{gather}
    [D_{\mu}, F^{\mn}] = 0
\end{gather}
until $\tau_{\rm hyd} = \qty{0.52}{\femto\meter}$.
At that time, the CYM stress-energy tensor, given by
\begin{gather}
    T^{\mn} = \Tr \left( -g^{\mu \gamma}g^{\nu \alpha}g^{\beta \delta}F_{\gamma \beta}F_{\alpha \delta} 
    + \frac{1}{4}g^{\mu\nu}g^{\gamma \beta}g^{\alpha \delta}F_{\gamma \alpha}F_{\beta \delta}\right)
\end{gather}
is constructed. It is symmetric and gauge invariant, and is the bridge that connects the 
pre-equilibrium dynamics of the glasma to the relativistic hydrodynamics of the QGP. We 
refer to~\cite{BLASCHKE2016192} for an in-depth discussion about the properties and 
development of the tensor. To find the energy density $\varepsilon$ and flow velocity 
$u^{\mu}$, we diagonalize $T^{\mn}$ and preserve the timelike eigenvalue. The flow 
velocity is normalized to $u_{\mu}u^{\mu} = 1$ throughout.

The shear-stress tensor $\pi^{\mn}$, which is needed to initialize viscous hydrodynamics, 
is given by
\begin{gather}
    \pi^{\mn} = T^{\mn}_{\mathrm{IPG}} - T^{\mn}_{\mathrm{ideal}} \\
    T^{\mn}_{\mathrm{ideal}} = (\varepsilon + P)u^{\mu}u^{\nu} - Pg^{\mn}
\end{gather}
In IP-Glasma, the pressure $P$ is given by ${\varepsilon}/{3}$ due to
the conformality of the classical gluon system.
The conformal nature of the pre-equilibrium phase,
originating from the pure gluon and classical features of the CGC, also justifies the
absence of a bulk pressure $\Pi$ at this stage. On the hydro side, the pressure is
dictated by the EoS at use, which in this study is HotQCD~\cite{Bazavov_2018}. An issue
arises when one realises that $P_{\mathrm{IPG}} = {\varepsilon}/{3}$ and
$P_{\mathrm{EoS}}(\varepsilon)$ may not match, leading to a discontinuity in our
transition to hydrodynamics. This issue is handled by initializing the QGP with a bulk
pressure $\Pi$, which, at this point, is required and given by the difference between the
IPG and EoS pressures, i.e.
\begin{gather}
    \Pi = P_{\mathrm{IPG}}(\varepsilon) - P_{\mathrm{EoS}}(\varepsilon) = \frac{\varepsilon}{3} - P_{\mathrm{EoS}}(\varepsilon)
\end{gather}
We therefore fully conserve energy and momentum in our transition to hydrodynamics, 
which allows for proper tracking and accounting of quantities such as ${dE}/{d\eta}$ 
from the initial state into the hydrodynamics phase.

\subsection{MUSIC\label{subsec:music}}

\begin{figure}
    \includegraphics[width=\linewidth]{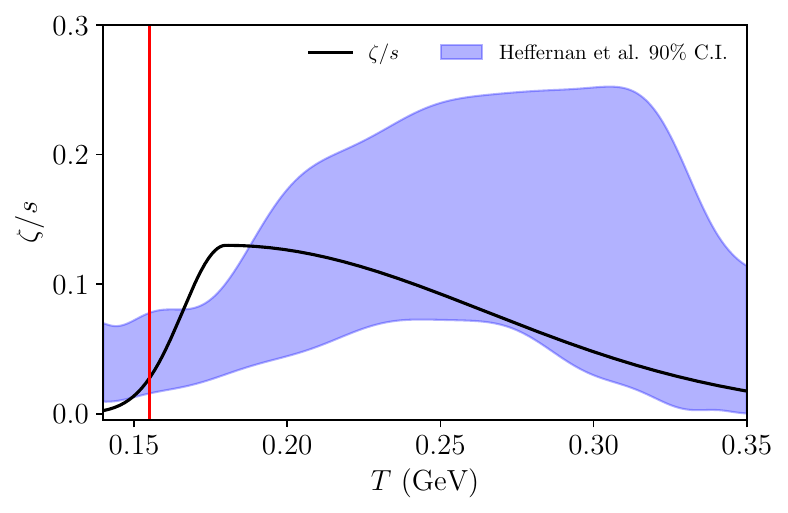}
    \caption{\label{fig:bulkparam} Bulk viscosity to entropy density ratio $\zeta / s$ as a 
    function of temperature. The red vertical line shows the freeze-out temperature ($T_{sw} = 155$ MeV) below
    which iS3D \& SMASH, our hadronic transport models, take over.}
\end{figure}

While the CGC is an effective \textit{field} theory, hydrodynamics is a long-wavelength
effective theory~\cite{Kolb:2003dz,Jeon_2015}. Both, however, have the same objective: to describe reality as
efficiently and accurately as possible, while providing pragmatic theories which allow for
the extraction of usable results. In the hydrodynamics phase, the evolution shifts from
more fundamental degrees of freedom such as the gluon fields 
to coarse-grained thermodynamic ensemble
averages, such as pressures and temperatures.

MUSIC~\cite{MUSIC_Schenke,Schenke:2010rr} is the numerical implementation of the 
following theoretical concepts. Formally, it is a second-order relativistic viscous 
hydrodynamics simulation. The fundamental quantity being evolved in hydrodynamics is 
the energy-momentum tensor
\begin{gather}
    T^{\mn}_{\mathrm{Hydro}} = T^{\mn}_{\mathrm{ideal}} + \Pi \left( u^\mu u^\nu - g^{\mn} \right) + \pi^{\mn}
\end{gather}
In second order viscous hydrodynamics,
the conservation laws
\begin{gather}
    \partial_{\mu} T^{\mn}_{\mathrm{Hydro}} = 0
\end{gather}
are supplemented by equations of motion for the 
the viscous tensor $\pi^{\mn}$ and the bulk pressure $\Pi$
\begin{gather}
    \dot{\Pi} = \frac{1}{\tau_{\Pi}}\left( -\Pi - 
    \zeta\Theta - \delta_{\Pi \Pi}\Pi \Theta
    \right) \label{eq:vis1} \\
   \dot{\pi}^{\langle\mn\rangle} = \frac{1}{\tau_{\pi}}\left( -\pi^{\mn} + 2 \eta \sigma^{\mn} -
    \delta_{\pi \pi}\pi^{\mn}\Theta \right) \label{eq:vis2}
\end{gather}
Here, $\Theta = \partial_\mu u^\mu$ is the scalar expansion rate and 
$\sigma^{\mn} = \nabla^{\langle\mu}u^{\nu\rangle}$ is the velocity shear tensor. 
The overdot represents the co-moving frame time derivative $u^\mu\partial_\mu$ and
the angular bracket around the indices indicate the transverse, symmetric and traceless
part of the tensor.
The coefficients $\delta_{\Pi \Pi}/\tau_{\Pi}$
and $\delta_{\pi \pi}/\tau_{\pi}$ are derived using the
14-moment approximation, whose values are ${2}/{3}$ and ${4}/{3}$ respectively,
according to~\cite{Denicol_2014}. The bulk viscosity $\zeta$ and the shear viscosity $\eta$ 
have the following form
\begin{gather}
    \frac{\zeta}{s}(T) = N\exp \left( -\frac{(T - T_p)^2}{B_{(L,G)}^2} \right) \label{eq:bulkvis} \\
    \frac{\eta}{s} = 0.136
\end{gather}
where $s$ is the entropy density and the energy unit is GeV.
In this study, following the Bayesian analysis
performed in~\cite{heffernan2023bayesian}, we have opted to use a constant shear viscosity $\eta /s$ and a temperature-dependent bulk viscosity $\zeta /s$.
Eq.\eqref{eq:bulkvis} shows the functional form of $\zeta /s$, where $N = 0.13$ and $T_p = \qty{0.18}{\giga\electronvolt}$. 
The width $B_{(L,G)}$ depends on whether the temperature of the QGP is smaller or 
larger than the bulk peak temperature $T_p$: if $T < T_p$, 
we have $B_L = \qty{0.02}{\giga\electronvolt}$, while $B_G = \qty{0.12}{\giga\electronvolt}$ is used for $T > T_p$. 
The form of the bulk is therefore that of an asymmetrical gaussian distribution, 
with features specifically chosen to avoid large bulk viscous corrections at freeze-out, as described in \cite{Schenke_2020}. 
Fig.~\ref{fig:bulkparam} shows Eq.~\eqref{eq:bulkvis} compared to the posterior 90\% confidence interval 
inferred in the above-mentionned Bayesian analysis~\cite{heffernan2023bayesian}. 
Our curve falls within the interval for most of the relevant temperature range. While its peak does fall outside of the confidence interval, 
our parametrization remains fully consistent with the findings of the analysis.
In Bayesian analysis,
a 90\% confidence interval provides a general region of space where a certain parameter should lie, 
while still allowing some part of the parameter curve to lie outside of it.
Given that the beam energies considered in this 
work are much smaller than those considered in~\cite{heffernan2023bayesian}, our results are more sensitive to low-temperature bulk features.
The bulk viscosity we have used reflects this sensitivity, while preserving consistency
with the Bayesian analysis in \cite{heffernan2023bayesian}.
With time, the QGP grows in volume while its temperature drops. At a certain temperature,
the QGP will `hadronize', i.e. turn into hadrons. The exact value of this temperature is
not sharply defined~\cite{Pandav_2022,yin2018qcd,Luo_2017}. In this
study, the switching temperature is set to $T_{\rm sw} = \qty{155}{\mega\electronvolt}$. Once
a specific fluid cell cools down to temperature $T_{\rm sw}$, its spatiotemporal state is saved.
Once every fluid cell has reached $T_{\rm sw}$, all of the 4-dimensional states are combined
to generate a constant-temperature hyper-surface, which terminates the hydrodynamic stage
of the simulation.

\subsection{iS3D \& SMASH\label{subsec:is3dsmash}}
The freeze-out hyper-surface generated by MUSIC is fed into iS3D~\cite{mcnelis2020particlization}, 
a particlization code which implements Cooper-Frye sampling~\cite{PhysRevD.10.186}, i.e.
\begin{gather}
    E\frac{dN_i}{d^3p} = \frac{d_i}{\left( 2\pi \right)^3} \int_{\Sigma} f_i(x,p)p_{\mu}d\sigma^{\mu}(x)
\end{gather}
where $\Sigma$ is an isothermal hyper-surface and $\sigma^{\mu}(x)$ is its normal vector,
$E\frac{dN_i}{d^3p}$ is the momentum spectrum of particle species $i$, $f_i(x,p)$ is its
phase-space distribution and $d_i$ is the degeneracy factor. To ensure a smooth transition
between hydrodynamics and particlization, the energy-momentum tensor $T^{\mn}$ must be
reproduced everywhere on the hyper-surface. Therefore,
\begin{gather}
    T^{\mn}_{\mathrm{kin}} = \sum_i d_i \int \frac{d^3p}{\left( 2\pi \right)^3
    E}p^{\mu}p^{\nu}\left( f_{{\rm eq},i}(x,p) + \delta f_i(x,p) \right)
\end{gather}
where $f_i(x,p)$ has been decomposed into an equilibrium distribution (which follows
either Bose-Einstein or Fermi-Dirac statistics depending on the species $i$) and an
out-of-equilibrium correction $\delta f_i$. This correction is necessary to account for
the viscous nature of our hydrodynamic evolution. Indeed, these lead to a medium which is
out-of-equilibrium at the time of sampling which in turn produces slight deviations from
the equilibrium distributions $f_{{\rm eq}, i}$. To match the nature of our shear stress
$\pi^{\mn}$ and bulk viscous pressure $\Pi$, we use the 14-moment $\delta f_i$
corrections. It is an expansion of $\delta f_i$ which is truncated at terms of first- and
second-order in momentum ($p^{\mu}$ and $p^{\mu}p^{\nu}$)~\cite{graddeltaf,
PhysRevC.80.054906, ISRAEL1979341}.

An important concept, which will be revisited in more detail in subsection
\ref{subsec:avg}, is the fact that the freeze-out hyper-surface from a single hydrodynamic 
event is oversampled hundreds of times~\cite{Schenke_2020,McDonald_2017,heffernan2023bayesian}.
Indeed, because the hyper-surface stems from a hydrodynamic treatment of the QGP, which
itself is an ensemble average, the sampling of particles will converge to the
hydrodynamic value of all observables (multiplicity, momenta, flow)
only once a sufficient
number of samplings are done. Therefore, a single IP-Glasma and MUSIC event, comprised of
a unique collision, impact parameter and nuclei configuration, can generate hundreds of
iS3D events, each consisting of its own particle list containing specific species and momenta.
In this work, we test two different ways of treating these oversampled events.
The oversampled events from a single hydrodynamic event
can be averaged over and treated as a single event, relating it to its initial
and hydrodynamic stages uniquely.
Alternatively, each oversampled event can be regarded as a distinct event fitting
the general prescription provided by the ensemble average hyper-surface. Choosing one method
over the other has tangible effects on computed observables, as will be evidenced in our
results.

Once the hadrons are sampled, they are evolved kinetically using SMASH
\cite{Weil_2016}, a hadronic cascade code. It implements inter-particle interactions and
scatterings, as well as resonance and decays
via coupled Boltzmann equations of most of the known hadrons.
We used SMASH Version 1.8 in this work.

\section{Methods and Observables\label{sec:obs}}
\subsection{Centrality\label{subsec:cent}}

Proper centrality selection is key to ensuring that the theoretical and computational
models of HICs are comparable to available experimental data. In this study, matching
multiplicities across centrality classes served as the sole calibration of our model: the
proportionality constant $C = 0.505$, described in section \ref{subsec:ipg},
was calibrated
to reproduce U+U multiplicity distributions at $\qty{193}{\giga\electronvolt}$ following the approach advocated in 
\cite{McDonald_2017} and described below.
This is the only calibration we made. All observables extracted thereafter used this
calibrated value.

When all events have gone through all stages of our framework, they are sorted by charged
particle multiplicity $dN_{\textrm{ch}} / d\eta$, then separated into a sufficiently many number of bins, all
of which contain the same number of events. The $C$ parameter is calibrated such that the 
average multiplicity of the most central bin matches that of experiment. Two experimental 
centrality bins and their respective multiplicities are then selected. These are the most 
central and the most peripheral centrality we would like to analyze (in this study, $0-1\%$ and
$27-28\%$ respectively). The experimental multiplicity ratios of these two bins
is then computed. The same is done with the first and last bins
from the simulations. While the ratio of the average multiplicities of our two selected
bins exceeds that of the corresponding experimental ratio, we drop the lowest-multiplicity
event from our consideration, recalculate the ratio, and keep repeating this until the
desired ratio is achieved, allowing us to reject a negligible number of events compared to
what would have to be rejected in a fully minimally-biased study. The calibration and
peripheral event dropping steps are sufficient to ensure that we reproduce available
charged hadron multiplicity curves (see Fig.~\ref{fig:Nchvcent} below),
which then allows for a thorough comparison to all
available observables.

Another important aspect of this study was the faithful emulation of Zero-Degree
Calorimeter (ZDC) binning to compare to data from STAR~\cite{Adamczyk_2015}. The ZDC is a
calorimeter that resides at $0^{\circ}$ from the beam direction.
It aims to
measure neutrons which were a part of the colliding nuclei, but did not 
participate in the collision.
Their lack of electric charge means that once they are free from the confines of
their nucleus, the collider's electromagnetic fields do not affect them, leading them to
follow straight paths directly into the calorimeters.

Experimentalists assess the centrality of a collision by counting the number of detected
neutrons: the more neutrons were found, the higher the chance that the collision was
peripheral. To emulate ZDCs within our framework, we calculate the total number of
participating nucleons from a given collision event and subtract it from the total number
of nucleons available to give us the number of spectator nucleons $S$, i.e.
\begin{gather}
    S = 2A - N_{\mathrm{Participants}}
\end{gather}
where $A = 238$ for a U+U collision. To obtain the number of neutrons out of the total 
number of spectator nucleons, we sample a binomial distribution
\begin{gather}
   P(N) = \binom{S}{N}\left( 1 - \frac{Z}{A} \right)^N \left( \frac{Z}{A} \right)^{S - N}
\end{gather}
where Z is the atomic number ($92$ for U) and we aim to sample $N$, the number of
neutrons, as done in~\cite{BBUUIPG}. We average $20$ samplings of this distribution per
event to reduce variability, giving us the number of spectator neutrons for each event.
This method does overlook some points, such as the fact that atomic nuclei have neutron
skins (an outer shell where only neutrons are found)~\cite{Liu_2022,
giacalone2023determination} which lead to higher probabilities of having spectator
neutrons than protons which aren't encapsulated within our simple binomial distribution.
These considerations, however, were outside of the scope of this study.

\subsection{Averaging\label{subsec:avg}}

In section \ref{subsec:is3dsmash}, we briefly discussed the concept of oversampling the hydrodynamical
hyper-surface. Given the ensemble average definition of this hyper-surface, sampling it
multiple times is an important part of ensuring that the final state observables
associated with an event converge to their hydrodynamic values. Previously
\cite{Schenke_2020,McDonald_2017,heffernan2023bayesian}, this oversampling procedure
(which, in this work, we will call `oversampled average') is followed by an averaging of
the oversampled events to create a single set of observables related to a single
hyper-surface. By design, this method 
smoothes out fluctuations and accentuates the 
features associated with a given hyper-surface.
If, however, each oversampled event is regarded
as an independent event and averaging is only performed
in a given centrality class, the effects of 
short-range fluctuations and
correlations will remain
(we will call this method `SMASH sub-event averaging').
However, events that came from the same hyper-surface event may not be fully independent.
Not all observables are sensitive to these differing averaging methods.
But, as will become evident in our results section,
many observables studied in this work {\em are} sensitive to the averaging method.

An ideal simulation of one experimental event 
would be the chain of one IP-Glasma, one hydro, and one SMASH event.
This, however, requires many orders of magnitude more computing
resources than are currently available. 
As such, it is important to analyze which observables are sensitive and for what reasons,
as we have done in this study.
For this purpose, we also compare the above two averaging methods 
to the `mixed events' in Section \ref{sec:res}, where
we group all oversampled events from a given centrality class, mix all of their
particles and create new `mixed' events. Only un-correlated
fluctuations should survive in these mixed events.

\subsection{Defining Selected Observables\label{subsec:misc}}

The following parts will briefly introduce and define a selection of the observables which
will figure in our results in section \ref{sec:res}. 

\subsubsection{Flow analysis\label{subsec:flow}}

The \textit{n}-th anisotropic flow coefficient $v_n$ is by now generally accepted as 
one of the primary evidence of QGP undergoing fluid-like behaviour in relativistic heavy-ion 
collisions. In this study, we will be interested in the 2- and 4-particle cumulants of various 
components of the flow harmonics. To start, we define the flow vector $Q_n$ for each event~\cite{Borghini_2001},
\begin{gather}
    Q_n = \sum^{N_{\mathrm{ch}}}_{j = 1} e^{in\phi_j}
\end{gather}
where $N_{\mathrm{ch}}$ is the event's multiplicity, $j$ runs over all of the particles of 
the event with transverse momentum restricted to $\qty{0.2}{GeV} < p_T < \qty{2.0}{GeV}$ to 
conform with the STAR acceptance window, and $\phi_j$ is the azimuthal angle of the 
$j^{\mathrm{th}}$ particle. Then, the $2^{\mathrm{nd}}$ order azimuthal correlation is given by
\begin{gather}
    \fluc{2} = \frac{\abs{Q_n}^2 - N_{\mathrm{ch}}}{N_{\mathrm{ch}}(N_{\mathrm{ch}} - 1)}
\end{gather}
while the $4^{\rm th}$ order azimuthal correlation is
\begin{multline}
    \fluc{4} = \frac{\abs{Q_n}^4 + \abs{Q_{2n}}^4 - 2\operatorname{Re}\left[ Q_{2n} Q_n^*Q_n^*\right]}{N_{\mathrm{ch}}(N_{\mathrm{ch}} - 1)(N_{\mathrm{ch}} - 2)(N_{\mathrm{ch}} - 3)} \\ 
   - 2\frac{2(N_{\mathrm{ch}} - 2)\cdot\abs{Q_n}^2 - N_{\mathrm{ch}}(N_{\mathrm{ch}} - 3)}{N_{\mathrm{ch}}(N_{\mathrm{ch}} - 1)(N_{\mathrm{ch}} - 2)(N_{\mathrm{ch}} - 3)}.
\end{multline}
where $Q_{2n}$ is to be understood as the flow vector associated with the \textit{2n}-th
harmonic if we are calculating the $4^{\mathrm{th}}$ order azimuthal correlation of the
\textit{n}-th harmonic (i.e. if $n = 2$, then $Q_{2n} = Q_4$). We then take an average of
these correlations over the entirety of events in their centrality class, which finally
allows us to compute the 2- and 4-particle cumulants, i.e.
\begin{gather}
    v_n\{2\} = \sqrt{\fluc{\fluc{2}}} \\
    v_n\{4\} = \sqrt[4]{-\left( \fluc{\fluc{4}} - 2 \cdot \fluc{\fluc{2}}^2 \right)}
\end{gather}
where $\fluc{\fluc{\cdot}}$ denotes $\fluc{\cdot}$ averaged over the given centrality.

Finally, the 2-particle scalar product $p_T$-differential flow is given by
\begin{gather}
    v_n\{2\}(p_T) = \frac{\operatorname{Re}(\fluc{Q_n^{\rm PI}(p_T) 
    \cdot (Q_n^{\rm ref})^{*}})}{\fluc{N_{\rm ch}^{\rm PI}(p_T)
    N_{\rm ch}^{\rm ref}}v_n^{\rm ref}\{2\}}
\end{gather}
where the superscript `PI' denotes the particle species of interest, while the superscript
`ref' denotes the reference flow vector. To avoid self-correlations being represented in
this observable, $Q_n^{\rm PI}$ is taken from the $| \eta | < 0.5$ rapidity window, while
the reference flow vector is taken from $0.5 < \eta < 2$.

For a more thorough treatment and discussion of these quantities, along with their 
respective errors, see~\cite{Bilandzic_2011}.

\subsubsection{Multi-particle Transverse Momentum Correlators\label{subsubsec:multi}}
We will be presenting results for 2- and 3-particle $p_T$ correlations, which are 
sometimes referred to as `variance' and `skewness' respectively, the difference being 
that the correlators do not consider self-correlations. The event-averaged 2-particle 
transverse momentum correlator is defined as 
\begin{gather}
    \fluc{\delta p \delta p} = \fluc{\frac{\sum_{i \neq j}(p_i - \fluc{p_T})(p_j -
     \fluc{p_T})}{N_{\mathrm{ch}}(N_{\mathrm{ch}} - 1)}}
\end{gather}
where the sum is over particles in a given event, the averaging is over the given
centrality class and $\fluc{p_T}$ denotes the average transverse momentum in the
centrality class being analyzed. The 3-particle version is
\begin{multline}
    \fluc{\delta p \delta p \delta p} = \\
    \fluc{\frac{\sum_{i \neq j \neq k}(p_i - \fluc{p_T})(p_j - \fluc{p_T})(p_k - 
    \fluc{p_T})}{N_{\mathrm{ch}}(N_{\mathrm{ch}} - 1)(N_{\mathrm{ch}} - 2)}}
\end{multline}
where $\fluc{p_T}$ denotes the average transverse momentum over the centrality 
class being analyzed. Implementing these formulae numerically as they are 
presented would be unwise, as they would run at least in $O(N_{\mathrm{events}} 
\cdot N_{\mathrm{ch}}^2)$ and $O(N_{\mathrm{events}} \cdot N_{\mathrm{ch}}^3)$ 
respectively. To avoid such computationally taxing and redundant computations, 
we have implemented a modified version of a framework presented by Giacalone et al. 
\cite{Giacalone_2021}. We start by defining $P_n$, the modified moments of the $p_T$ 
distributions,
\begin{gather}
    P_n = \sum_i^{N_{\mathrm{ch}}} (p_i - \fluc{p_T})^n
\end{gather}
where $p_i$ is the transverse momentum of the $i^{\mathrm{th}}$ particle in the 
given event. Then one can easily show
\begin{gather}
    \sum_{i \neq j}(p_i - \fluc{p_T})(p_j - \fluc{p_T}) = (P_1)^2 - P_2
\end{gather}
\begin{multline}
    \sum_{i \neq j \neq k}(p_i - \fluc{p_T})(p_j - \fluc{p_T})(p_k - \fluc{p_T}) = \\
    (P_1)^3 - 3P_2P_1 + 2P_3 
\end{multline}
allowing for the following redefinitions of the 2- and 3-particle transverse momentum correlators 
\begin{gather}
    \fluc{\delta p \delta p} = \fluc{\frac{(P_1)^2 - P_2}{N_{\mathrm{ch}}(N_{\mathrm{ch}} - 1)}}\\
    \fluc{\delta p \delta p \delta p} = \fluc{\frac{(P_1)^3 - 3P_2P_1 + 2P_3 }{N_{\mathrm{ch}}(N_{\mathrm{ch}} - 1)(N_{\mathrm{ch}} - 2)}} 
\end{gather}
All of the modified moments are computable in linear time on an event-by-event basis, 
greatly reducing the computational stress required to extract such observables from 
large particle lists. Also, since both quantities have been reduced to a single term 
compared to the moments presented in~\cite{Giacalone_2021}, the calculation of their 
respective errors has been greatly simplified.
\subsubsection{Transverse-momentum-flow correlations\label{subsec:rho}}
The final selected observable integrates both the flow harmonics and the 2-particle 
transverse momentum correlator. It is a correlator between 2-particle flow harmonics 
and average transverse momentum which was developed in~\cite{Bo_ek_2016}. It is defined as 
\begin{gather}
    \rho(v_n\{2\}^2,\fluc{p_T}) = \frac{\operatorname{cov}(v_n\{2\}^2, 
    \fluc{p_T})}{\sqrt{\operatorname{var}\left( v_n^2 \right)\cdot
    \fluc{\delta p \delta p}}} \label{eq:correlator}
\end{gather}
where
\begin{multline}
    \operatorname{cov}(v_n\{2\}^2, \fluc{p_T}) = \\
    \left\langle \frac{\abs{Q_n}^2 - N_{\mathrm{ch}}}{N_{\mathrm{ch}}(N_{\mathrm{ch}} - 1)} 
    \cdot \left(\frac{\sum_{i = 1}^{N_{\mathrm{ch}}} p_i}{N_{\mathrm{ch}}} - \fluc{p_T}\right) \right\rangle \label{eq:covar}
\end{multline}
and 
\begin{gather}
    \operatorname{var}\left( v_n^2 \right) = v_n\{2\}^4 - v_n\{4\}^4
\end{gather}
This correlator will be important in highlighting specific properties of central 
collisions of deformed nuclei. Indeed, it should show marked differences when 
compared to results from collisions of spherically symmetric nuclei.

\section{Results and Discussion\label{sec:res}}

Our results section will be divided into two subsections; the first contains 
comparisons of our model to existing data from two RHIC detectors (STAR and 
PHENIX) for U+U collisions at $\sqrt{s_{NN}} = \qty{193}{\giga\electronvolt}$ when 
available and Au+Au collisions at $\sqrt{s_{NN}} = \qty{200}{\giga\electronvolt}$ otherwise,
 while the second will focus on predictions of our model regarding multiparticle 
 correlations and transverse-momentum-flow correlations in U+U collisions.

\subsection{Descriptions of Existing Data}

\subsubsection{Charged Hadron Multiplicity}

We begin by verifying that our model can reproduce charged hadron multiplicity at
midrapidity.
As mentioned in section \ref{subsec:cent}, this observable serves as the sole
calibration tool for the proportionality constant between the saturation scale $Q_s$ and
colour charge fluctuations $\mu_A$. The proportionality constant was calibrated using U+U experimental
data and the Prev $^{238}$U parametrization Tab.~\ref{tab:params}.

\begin{figure}
    \includegraphics[width=\linewidth]{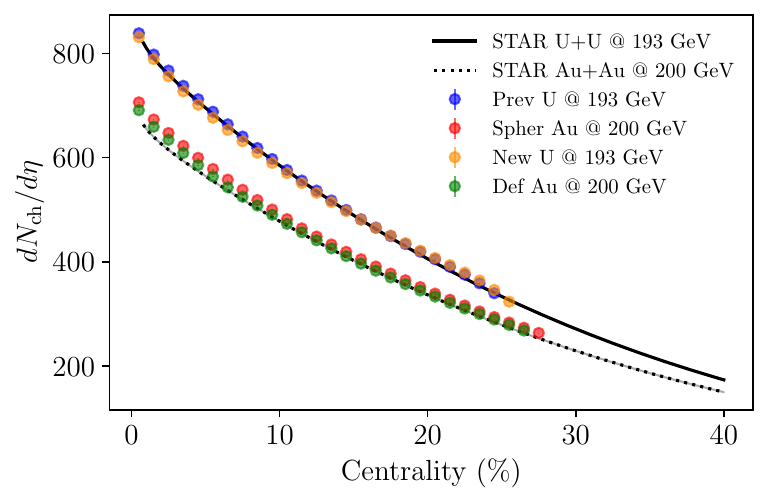}
    \caption{\label{fig:Nchvcent} Charged hadron multiplicities in $|\eta| < 0.5$ as 
    a function of centrality in our model, compared to results for 
    $\qty{193}{\giga\electronvolt}$ U+U and $\qty{200}{\giga\electronvolt}$ 
    Au+Au collisions at STAR~\cite{Adamczyk_2015}.}
\end{figure}

In Fig.~\ref{fig:Nchvcent}, we show the number of charged particles per unit
pseudo-rapidity $dN_{\rm ch}/d\eta$ from our model compared to data from STAR at
$\qty{193}{\giga\electronvolt}$ (U+U) and $\qty{200}{\giga\electronvolt}$ (Au+Au). The
experimental data stems from a parametrization undertaken in~\cite{Adamczyk_2015}. Our
model's agreement with the U+U and Au+Au data is excellent throughout, with the 
Au data showing a slight overestimate of the multiplicity in central collisions.
Comparing both systems' curves, we find that their features are similar beyond 
the fact that U collisions yield more hadrons, which is sensible given their 
larger nucleonic content (and, consequently, their larger total collision energy).

For the U parametrizations, the most peripheral
point, at $25-26\%$ centrality, dips slightly compared to the rest of our curve. This is due
to our centrality selection procedure outlined in section \ref{subsec:cent}. Indeed, our
procedure is bound to allow for events which are `too' peripheral to be found in our most
peripheral bin, given that we reject events based on multiplicity (and, therefore,
peripherality) until the multiplicity ratio matches that of the experiment. 
The Au parametrizations, on the other hand, do not feature this same dip, as 
the size of Au nuclei (spherical or deformed) is smaller than that of U nuclei. Therefore, given the 
impact parameter range used in this study ($\qty{0}{\femto\meter} \leq b \leq \qty{8}{\femto\meter}$), 
our raw Au data actually extended beyond the $25\%$ maximum centrality of our U calculation. We elected 
to apply our centrality selection procedure with a similar target range as our U data, as more peripheral 
data would not be of use in our analysis. This meant that our centrality selection cut a considerable amount 
of Au events off, but it also meant that it was more robust, as this centrality point's associated impact 
parameter interval was well-within our chosen impact parameter range. It is clear that, for both collision systems,
the choice of parametrization has little effect on multiplicity across the entire range, 
with the largest difference being in central ($0 - 5\%$ centrality) collisions of Au nuclei. 
This overestimate, both in scale and in position, is 
not surprising given that differences between parametrizations are exacerbated in 
central collisions and that our calibration was U-specific.

Our relatively narrow centrality window ($\sim 0 - 25 \%$) is due to a conscious 
choice and focus: the differences between collisions of deformed nuclei and 
collisions of spherical nuclei are most prominent in central collisions. We 
sought to limit our scope to more central collisions to generate sufficient 
statistics in the region of interest without using computational resources 
to simulate more peripheral events where the differences are not particularly noticeable. 

\begin{figure}
    \includegraphics[width=\linewidth]{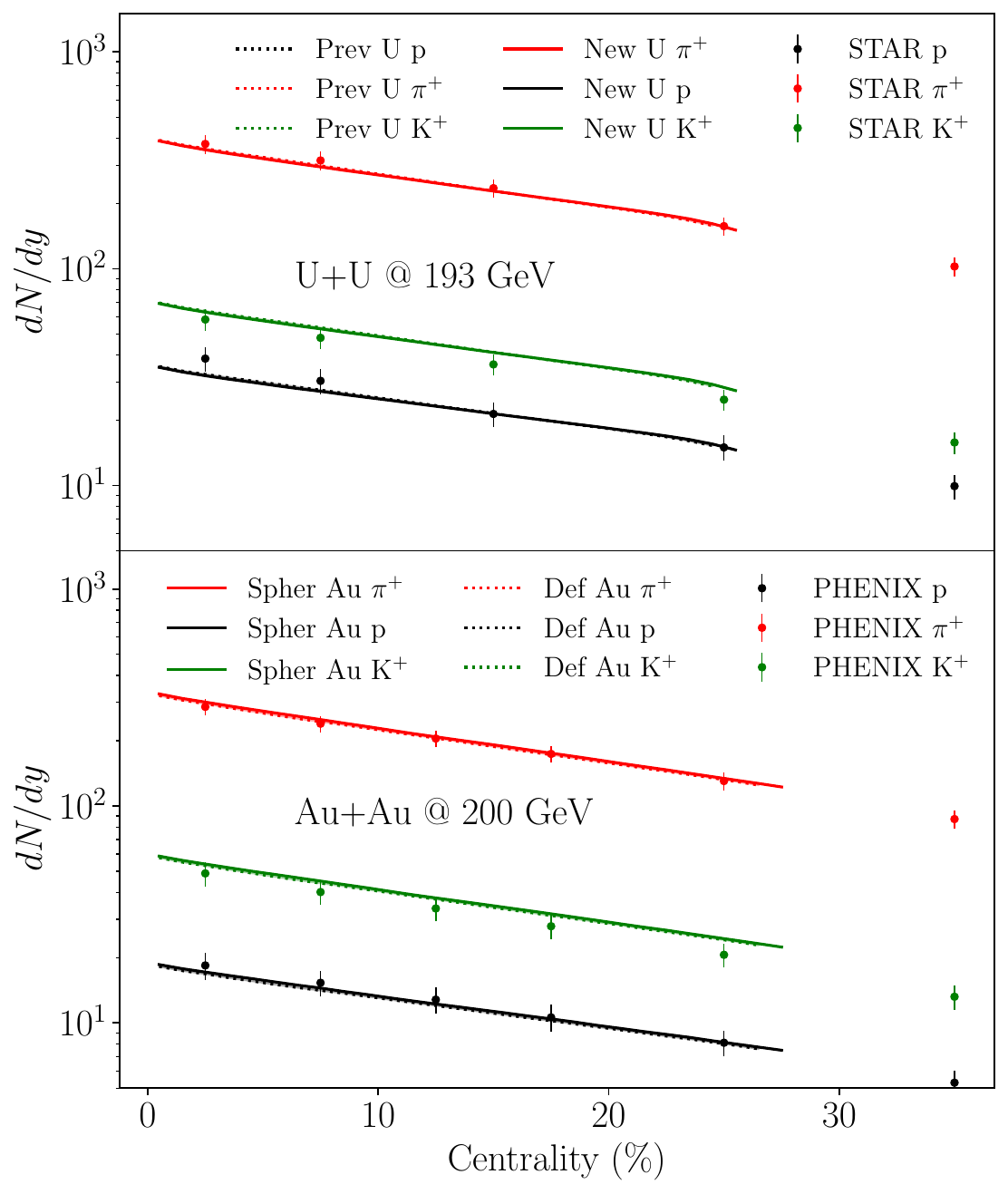}
    \caption{\label{fig:identifiedN} Identified particle multiplicity in $|y| < 0.5$ 
    as a function of centrality in our model. (\textbf{Top}) Two U configurations compared to results for 
    $\qty{193}{\giga\electronvolt}$ U+U collisions at STAR~\cite{Abdallah_2023}. (\textbf{Bottom}) Two Au 
    configurations compared to results for 
    $\qty{200}{\giga\electronvolt}$ Au+Au collisions at PHENIX~\cite{Adler_2004}.}
\end{figure}

Fig.~\ref{fig:identifiedN} shows identified particle yields as a function of centrality
compared to data from STAR~\cite{Abdallah_2023} and PHENIX~\cite{Adler_2004}. All species considered, our calculation shows great agreement
across the entire centrality window for all configurations and collision systems.
To mirror experimental procedures used at STAR~\cite{Abdallah_2023}, the pion yields
were corrected for feed-downs, while proton yields were not.
When compared to PHENIX data~\cite{Adler_2004}, both proton and pion yields were 
corrected for feed-downs.
It is important to note here that our model does not include
a baryon chemical potential. At this collision energy, $\mu_B$ is small but non-zero. 
Given that charged particle multiplicities were quasi-identical across different configurations,
it is not surprising that identified particle yields behave similarly.

\begin{figure}
    \includegraphics[width=\linewidth]{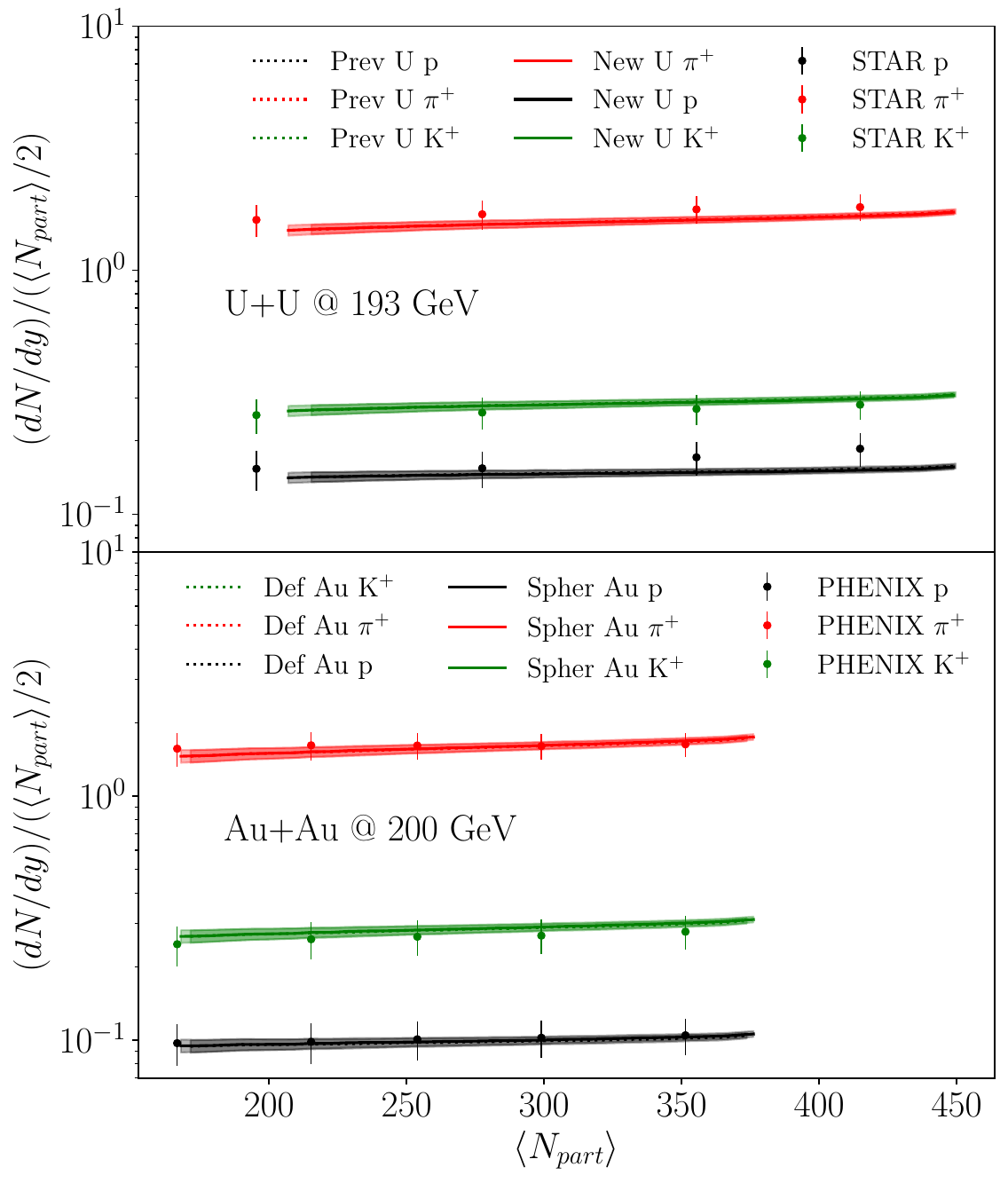}
    \caption{\label{fig:identifiedN_scaled} Identified particle multiplicity in $|y| <
    0.5$ scaled by average number of participant nucleon pairs in the centrality class
    $\fluc{N_{\rm part}}/2$ as a function of $\fluc{N_{\rm part}}$ in our model. 
    (\textbf{Top}) Two U configurations compared to results for 
    $\qty{193}{\giga\electronvolt}$ U+U collisions at STAR~\cite{Abdallah_2023}. (\textbf{Bottom}) Two Au 
    configurations compared to results for 
    $\qty{200}{\giga\electronvolt}$ Au+Au collisions at PHENIX~\cite{Adler_2004}.}
\end{figure}

Fig.~\ref{fig:identifiedN_scaled} shows identified particle yields scaled by the
average number of participant nucleon pairs in a given centrality class
$\fluc{N_{\rm part}}/2$ as a function of the number of participants. This set of results is dependent on the results shown in Fig.~\ref{fig:identifiedN}, as the number of
participant nucleons and centrality are highly correlated. However, this specific
observable looks to identify where particle production comes from at a given centrality,
and how it progresses across the spectrum. Because it increases with the average number of
participant nucleons, we determine that particle production is guided by a combination of
soft and hard processes that scale differently with $N_{\rm part}$. Here again, both configurations
for both collision systems give quasi-identical results, which entails that particle production mechanisms
are not tied to initial nuclear configurations.

\subsubsection{Average Transverse Momentum}

\begin{figure}
    \includegraphics[width=\linewidth]{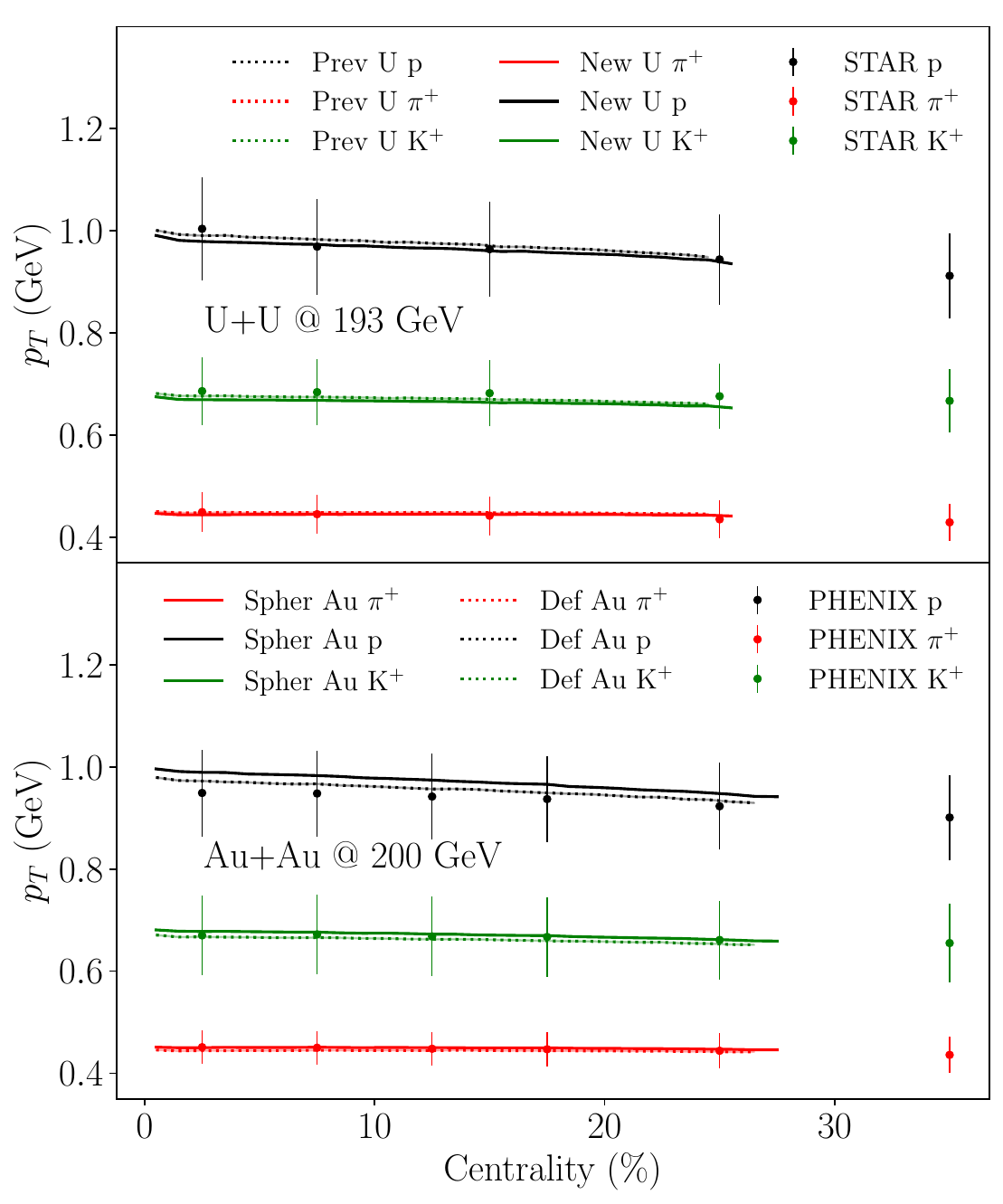}
    \caption{\label{fig:identifiedpT} Identified particle mean transverse 
    momentum $\fluc{p_T}$ in $|y| < 0.5$ as a function of centrality in our model. 
    (\textbf{Top}) Two U configurations compared to results for 
    $\qty{193}{\giga\electronvolt}$ U+U collisions at STAR~\cite{Abdallah_2023}. (\textbf{Bottom}) Two Au 
    configurations compared to results for 
    $\qty{200}{\giga\electronvolt}$ Au+Au collisions at PHENIX~\cite{Adler_2004}.}
\end{figure}

\begin{figure*}
    \includegraphics[width=\linewidth]{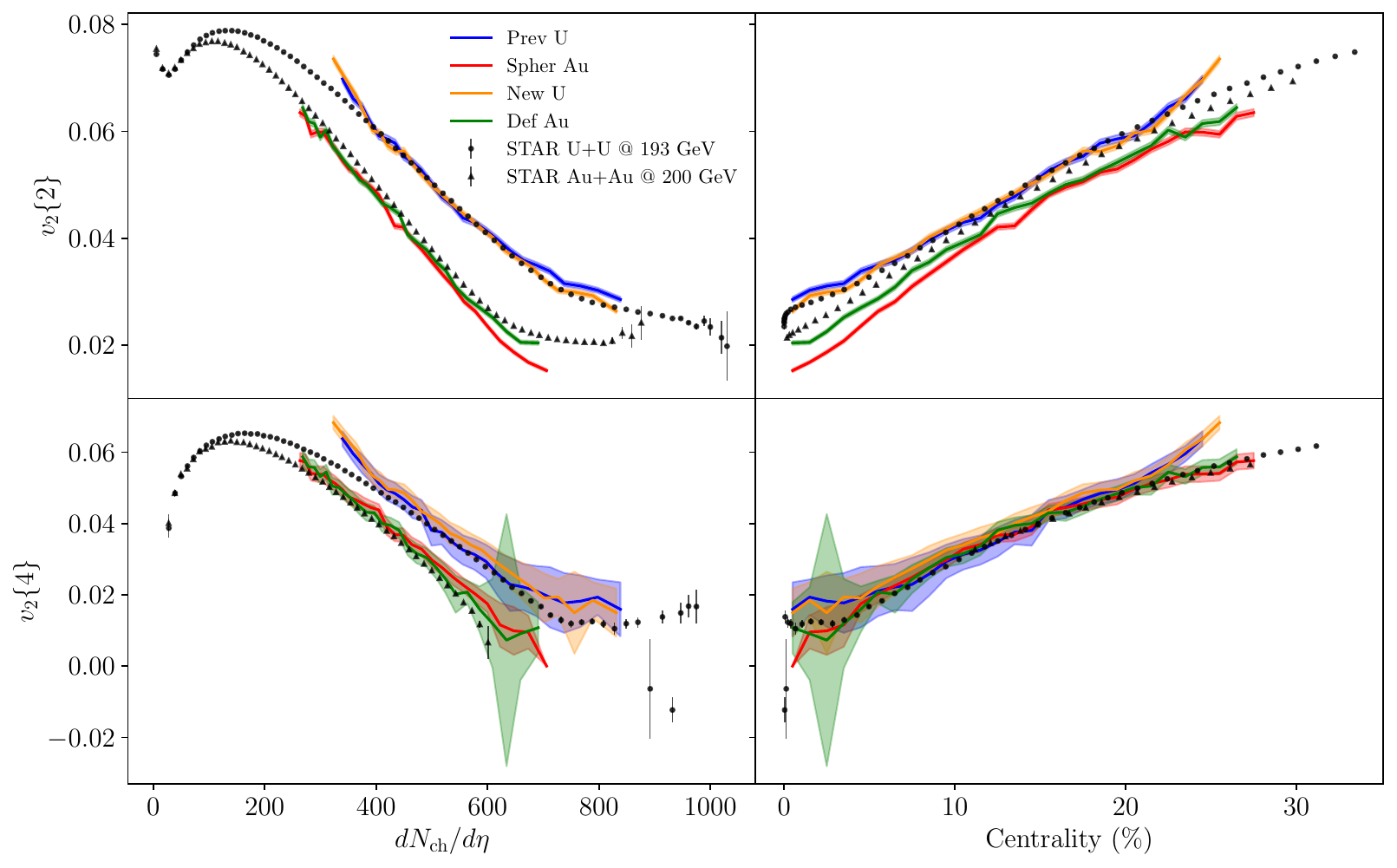}
    \caption{\label{fig:chargedv22} Two- and four-particle cumulants of 
    elliptic flow ($v_2\{2\}$ and $v_2\{4\}$) as functions of (\textbf{left}) 
    charged particle multiplicity and (\textbf{right}) centrality, compared to 
    results for $\qty{193}{\giga\electronvolt}$ U+U and $\qty{200}{\giga\electronvolt}$ Au+Au collisions at STAR 
   ~\cite{Adamczyk_2015, Abelev_2009}. The shaded bands represent statistical errors.}
\end{figure*}

We now move to the average transverse momentum of identified particles. 
Fig.~\ref{fig:identifiedpT} shows identified particle mean transverse 
momentum $\fluc{p_T}$ as a function of centrality. Here, while different configurations
do still give similar results, a more clear separation can be seen. This is mostly driven by the differences
in size of the various configurations. Indeed, a smaller nucleus leads to (generally) smaller
overlap areas in the transverse plane at equal collision energy, which in turn leads to higher energy densities and larger transverse momenta (again, at fixed collision energy). It is 
important to note here that size doesn't exclusively refer to the unmodified radius $R_0$ 
listed in Tab.~\ref{tab:params}, but rather to the combination of all Woods-Saxon parameters, 
as what we are really probing for is the shape of the overlap area between the nuclei, which 
is affected by all WS parameters simultaneously.
The respective masses of the identified particles are responsible for the ordering. Again, 
as with yields, kaons, pions and protons all show excellent agreement with the experimental data. 
It is important to note that given our use of hydrodynamics, $\fluc{p_T}$ puts strong 
constraints on the $p_T$ spectrum, which entails that a good agreement with 
$\fluc{p_T}$ is equivalent to reproducing the spectrum~\cite{Pratt_2015}. Therefore, 
given our excellent agreement with experimental data, the shear and bulk viscosity 
parametrizations, along with our switching temperature $T_{\rm sw}$ seem appropriate. 

\subsubsection{Anisotropic Flow \label{subsub:ani}}

We now shift our attention to anisotropic flow, focusing on integrated elliptic and triangular flows.
Fig.~\ref{fig:chargedv22} shows the two- and four-particle cumulants of elliptic 
flow ($v_2\{2\}$ and $v_2\{4\}$) as functions of charged hadron multiplicity and 
centrality. We see that our model reproduces both observables very well. All of these results were 
calculated using the typical oversampled averaging method. At smaller 
multiplicities (or more peripheral collisions), our model overestimates U elliptic flow. 
This is due to the same effect apparent in Fig.~\ref{fig:Nchvcent}, namely that our peripheral 
centrality class may contain more peripheral events which weren't rejected by our centrality 
selection process. It should be emphasized that no additional adjustments of parameters were made 
to produce our results. It should also be noted that the Au data does not have the same overestimating feature
in peripheral collisions, further proving that this is an artifact of our centrality selection process.

Looking at the two U configurations, one finds a slight (but marked) improvement going
from Prev $^{238}$U 
to New $^{238}$U. The slight decrease in $\beta_2^0$ leads to a general decrease in the two-particle 
cumulant of elliptic flow in central ($\sim 0 - 10\%$) collisions. The four-particle cumulant, on the other hand, seems
unaffected by the change, if only because the error bars are too large to draw conclusions.

The Au configurations show a marked split in central collisions, once again driven by the
inclusion (and exclusion) of $\beta_2^0$. In the most central collisions, the spherical parametrization
underestimates $v_2\{2\}$ by a wide margin, leading us to conclude that $^{197}$Au is deformed.
While the deformed $^{197}$Au configuration does mirror the experimental data well, it consistently underestimates
the experimental values throughout the multiplicity and centrality ranges. This underestimation, while small, does
open the door to future analyses using slightly larger values of $\beta_2^0$ and $\beta_2^2$.
Also interestingly, the four-particle cumulant shows absolutely no sensitivity to the changes in parametrization 
(from perfectly spherical to deformed), which, combined to similar observations for the two U configurations,
indicates that this cumulant is mainly driven by either higher-order deformation parameters or fluctuations.

\begin{figure}[t!]
    \includegraphics[width=\linewidth]{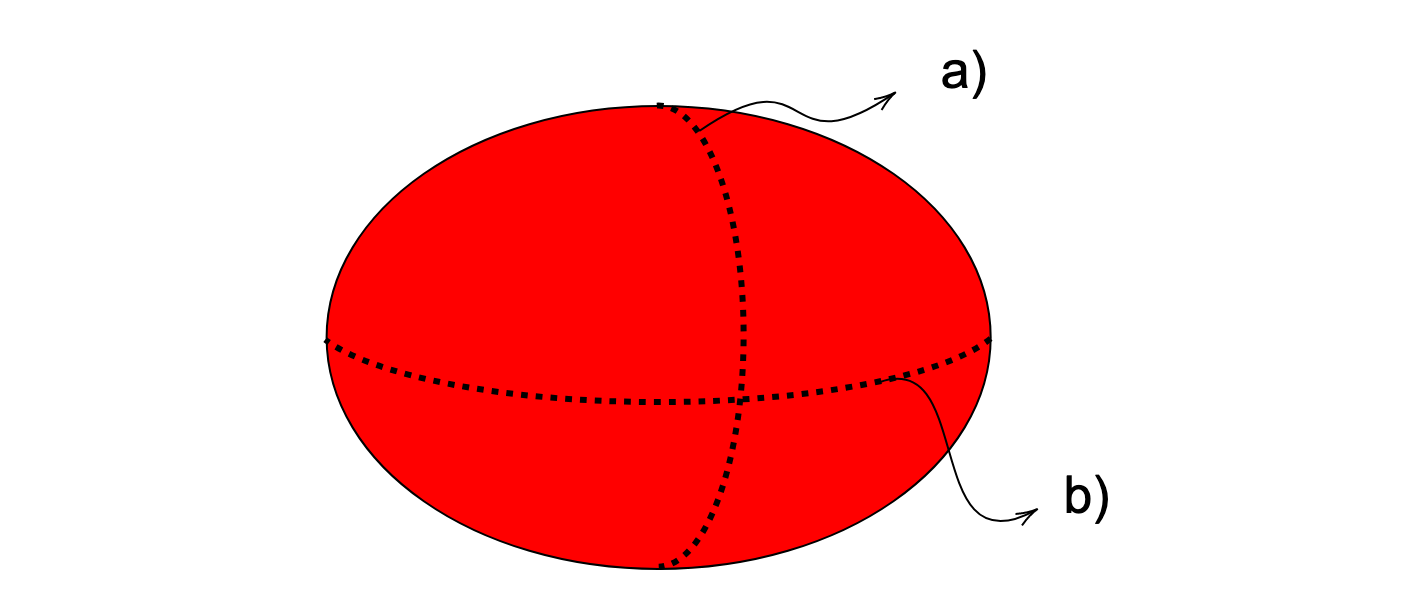}
    \caption{\label{fig:longshort} Schematic representation of the asymmetry 
    between a) the short-axis and b) the long-axis directions. In the short 
    axis direction, the nucleus is not deformed (has constant R in WS parametrization).}
\end{figure}
\begin{figure}[t!]
    \includegraphics[width=\linewidth]{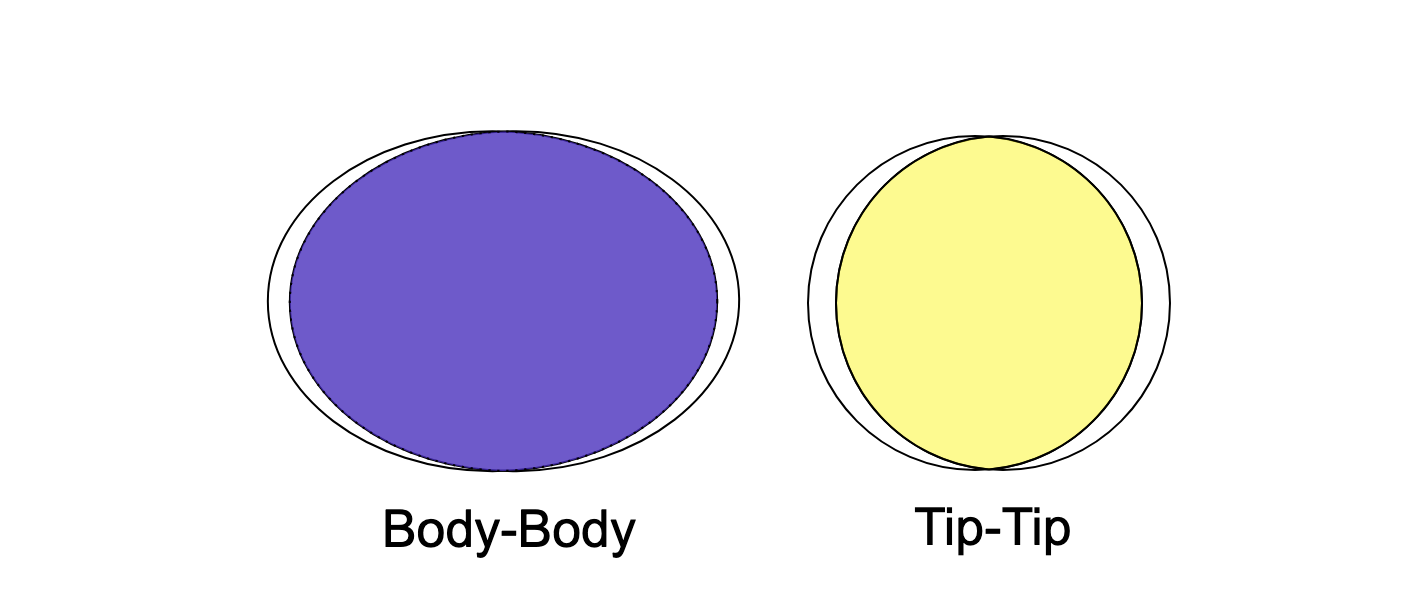}
    \caption{\label{fig:colltypes} Schematic representation of the difference 
    in nuclei overlap area between body-body and tip-tip collisions. The former 
    presents large eccentricity while the latter presents near-zero eccentricity.}
\end{figure}

One may notice that our model does not reach as high in multiplicity as the experimental 
data does. This is because the experimental points beyond $dN_{\textrm{ch}} / d\eta \approx 850$ 
constitute ultra-ultra-central collisions. That is, they are in the top $0.01\%$ of events 
registered at STAR; these experimental events are rare and would require a much larger number 
of runs on our part to reproduce. Nevertheless, the right side panels show that, in terms of 
centrality, we are essentially reproducing the entire spectrum of central events. 

These right-side panels also serve to compare Au and U data more accurately. We see that
experimental $v_2\{2\}$'s are extremely similar throughout the centrality spectrum, except
in the very central collisions ($0-5\%$). There, U data shows a marked increase before
coming back to Au levels between $0-1\%$. This is explained by the considerably deformed geometry of U.
In central collisions of spherically symmetric nuclei, only one overlap shape is generated
in the transverse plane, namely a circle, which has small (or 0) eccentricity. This in
turn leads to a small elliptic flow in the final state. In central collisions of largely deformed
nuclei such as U, many different transverse cross-sections are possible. Indeed, looking
at Figs.~\ref{fig:longshort} and \ref{fig:colltypes} we see that a nucleus travelling with
its long axis parallel to the beam direction will have a circular cross-section in the
transverse plane. Similarly, if its short axis is parallel to the beam direction, it will
have an elliptical transverse cross-section. Therefore, in collisions of deformed nuclei
with near-zero impact parameters, we can have both extremely eccentric (body-body) and
circular (tip-tip) cross-sections. The body-body collisions will have a smaller energy
density (due to their larger transverse overlap area) than their tip-tip counterparts,
which will in turn lead to slightly smaller multiplicities. Therefore, the marked increase
in $v_2\{2\}$ in $1-3 \%$ centrality is due to body-body collisions, and its sharp
decrease in ultra-central ($0-1 \%$) collisions is due to tip-tip collisions. 

These panels also serve as an indication that, while $^{197}$Au may be deformed, its deformation is 
fairly small compared to that of $^{238}$U, as the specific features described above are much less
prominent for $^{197}$Au than for $^{238}$U. They also further support the idea that $v_2\{4\}$ is driven
by fluctuations, as both the $^{197}$Au and $^{238}$U experimental results overlap throughout.

\begin{figure}[t!]
    \includegraphics[width=\linewidth]{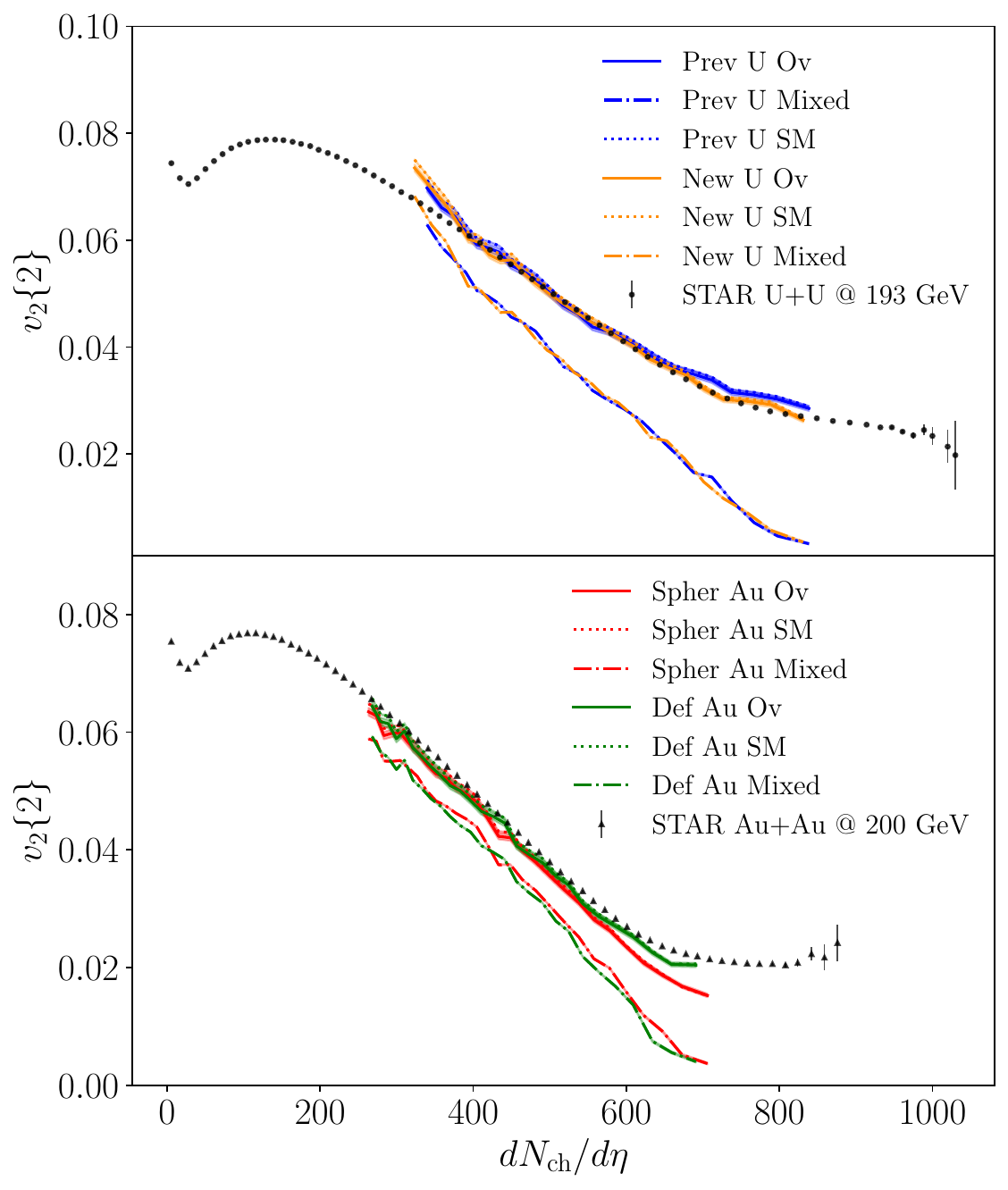}
    \caption{\label{fig:chargedv22mixed} Two-particle cumulant of elliptic 
    flow ($v_2\{2\}$) as functions of charged particle multiplicity for our two systems and four configurations, emphasizing 
    the addition of Mixed Event and SMASH sub-event average curves. (\textbf{Top}) Two U configurations compared to results for 
    $\qty{193}{\giga\electronvolt}$ U+U collisions at STAR~\cite{Abdallah_2023}. (\textbf{Bottom}) Two Au 
    configurations compared to results for 
    $\qty{200}{\giga\electronvolt}$ Au+Au collisions at STAR~\cite{Abelev_2009}. Here, \textit{SM} stands for SMASH sub-event average while \textit{Ov} stands for oversampled average.}
\end{figure}

Fig.~\ref{fig:chargedv22mixed} introduces an alternative averaging method described in section
\ref{subsec:avg}, namely the SMASH sub-event averaging. This sub-event averaging method
leads to the expression of short-range correlations which are usually suppressed by
oversampled averaging. We note that, for our two U configurations,
 this method leads to a slight overestimation of
$v_2\{2\}$ across the entire centrality range. For Au, we observe the same
behavior in moving from oversampled averaging to SMASH sub-event averaging, namely a small 
increase in $v_2\{2\}$. In all cases, however, both methods seem rather similar, indicating that 
short-range correlations and fluctuations do not play a significant role in
elliptic flow. As it stands, neither the
oversampling averaging nor the sub-event averaging faithfully follows what takes place in
a real heavy-ion collision because hydrodynamics is an inherently coarse-grained theory of
ensemble-averaged quantities. As such, one should regard the results of the two different
averaging procedures as a part of theoretical uncertainty ($\sim 5\,\%$).

Fig.~\ref{fig:chargedv22mixed} also sees the addition of a mixed event curve which was left out 
of Fig.~\ref{fig:chargedv22} for clarity. By mixing the events, only the average 
effect of the collision geometry should survive, whereas the effects due to the deformation
 will be washed out. This is indeed what can be observed in Fig.~\ref{fig:chargedv22mixed}:
 it clearly shows that deformation effects are crucial in understanding flow in U+U and Au+Au collisions. 

\begin{figure}[t!]
    \includegraphics[width=\linewidth]{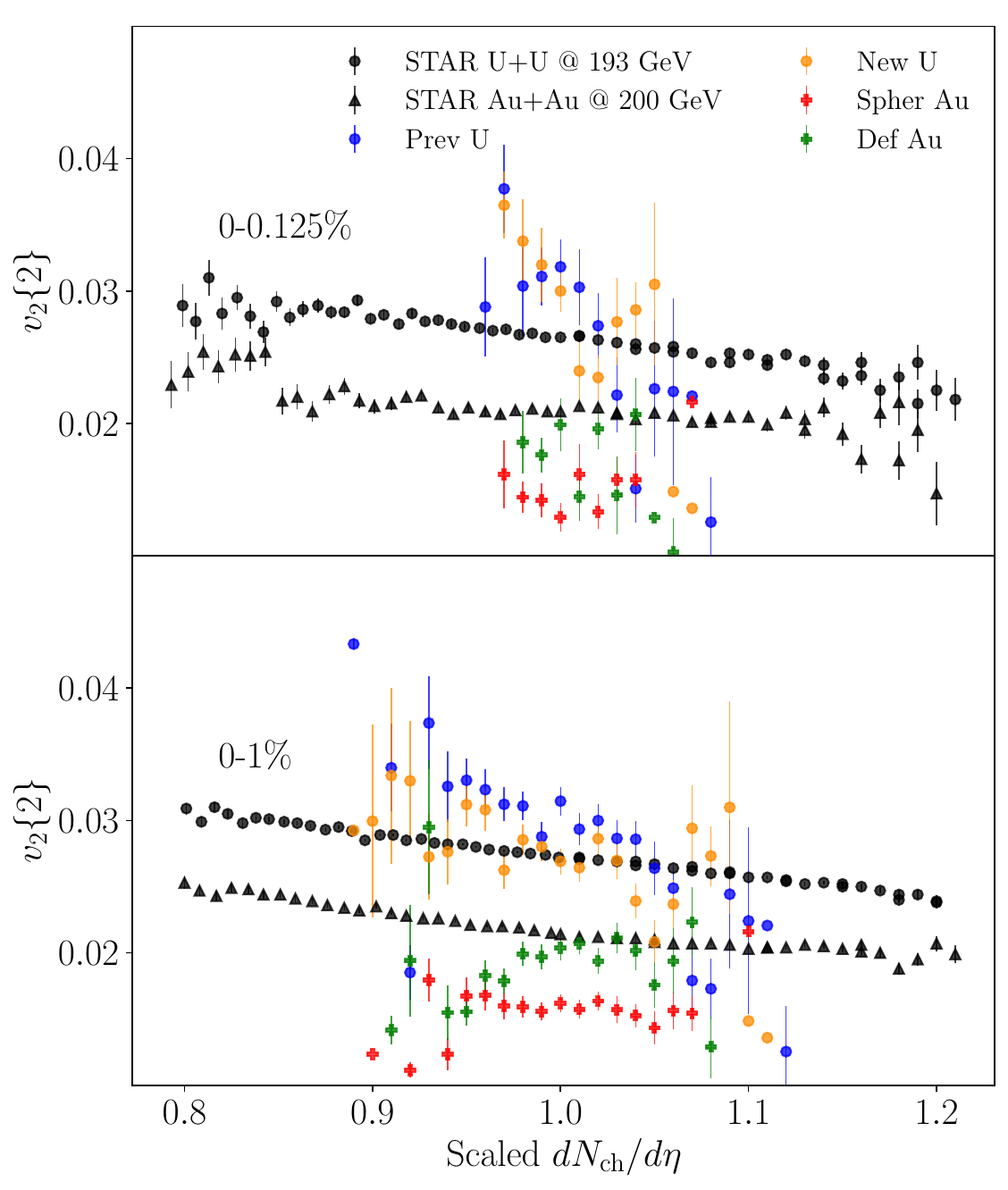}
    \caption{\label{fig:scaled} Two-particle cumulant of elliptic flow ($v_2\{2\}$) as 
    functions of scaled charged particle multiplicity for (\textbf{top}) $0-0.125 \%$ 
    and (\textbf{bottom}) $0-1 \%$ most cetral events, compared to results for 
    $\qty{193}{\giga\electronvolt}$ U+U and $\qty{200}{\giga\electronvolt}$ Au+Au collisions at STAR~\cite{Adamczyk_2015}.}
\end{figure}

Fig.~\ref{fig:scaled} shows the two-particle cumulant of elliptic flow as a function of
scaled multiplicity in two ultra-central ZDC bins: $0-0.125 \%$ and $0-1\%$. These events
were selected based on their respective number of sampled spectator neutrons, as described
in section \ref{subsec:cent}. For all configurations and collision
systems, the most central bin suffers from a small number of events,
which in turn affects statistics. Nevertheless, the scale of the experimental data is
reproduced. The Spher $^{197}$Au configuration underestimates the elliptic flow in both bins, 
which is expected considering Fig.~\ref{fig:chargedv22}, namely that this configuration was
 inappropriate for central collisions of Au nuclei.
In the broader ultra-central bin, our model reproduces the general shape and
trend of experimental data. For our two U configurations, we find New $^{238}$U to reproduce the
experimental data more faithfully, in line with what we've been able to conclude from Fig.~\ref{fig:chargedv22}. The
Def $^{197}$Au configuration performs decently in this broader centrality window. However, it considerably
underestimates elliptic flow at smaller multiplicities. This is most probably due to relatively
low statistics, but will require further investigation. 

\begin{figure}
    \includegraphics[width=\linewidth]{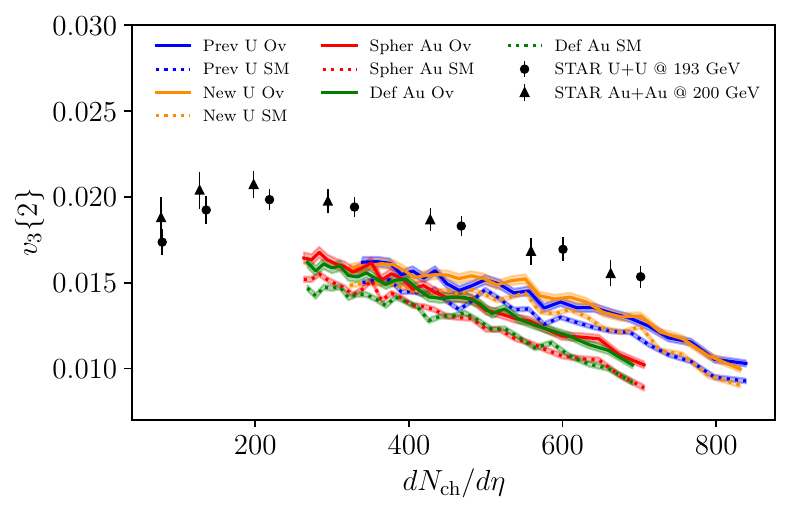}
    \caption{\label{fig:v32} Two-particle cumulant of triangular flow ($v_3\{2\}$) as a 
    function of charged particle multiplicity, compared to results for 
    $\qty{193}{\giga\electronvolt}$ U+U and $\qty{200}{\giga\electronvolt}$ Au+Au collisions at STAR~\cite{Adam_2019}. Here, \textit{SM} stands for SMASH sub-event average while \textit{Ov} stands for oversampled average.}
\end{figure}

Fig.~\ref{fig:v32} shows the two-particle cumulant of triangular flow $v_3\{2\}$ as a
function of charged particle multiplicity, which is a fluctuation-driven observable.
Looking at Fig.~\ref{fig:v32}, we see that experimental data for U and Au are practically
overlapping in central collisions, confirming that initial global geometry plays little
role in this observable. Our model underestimates triangular flow across our chosen range, 
collision systems and configurations, which indicates that it underestimates initial state fluctuations. 
This could potentially be mended by the addition of sub-nucleonic degrees of freedom (i.e.~valence quark
configurations) in our initial construction of the nuclear thickness function $T_{A}(\boldsymbol{x})$, such as
those described in~\cite{Kumar_2022}. Interestingly, we also see that our SMASH sub-event average
curves are consistently lower than their oversampled average counterparts. This result implies that our current 
exclusion of sub-nucleonic degrees of freedom leads to short-range fluctuations being noticeably smaller than longer-
range ones.

\begin{figure}
    \includegraphics[width=\linewidth]{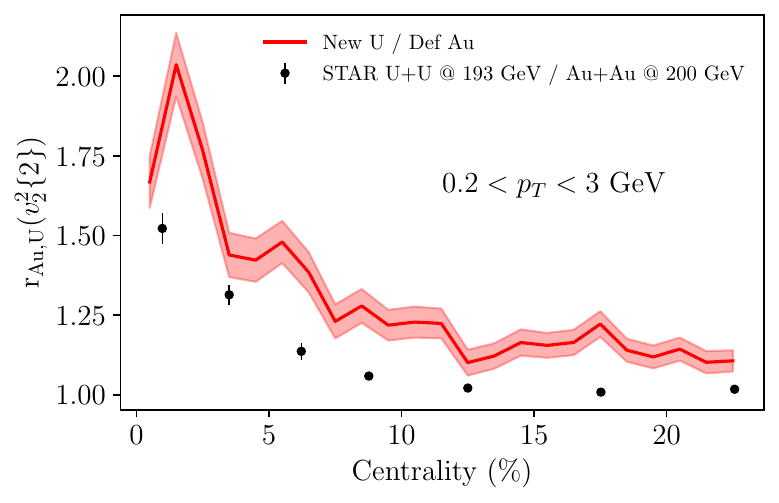}
    \caption{\label{fig:ratio} Ratio of mean squared elliptic flow $v_2\{2\}^2$ for our two best-fit
    parametrizations, New U and Def Au (c.f. Tab.~\ref{tab:params}), compared to 
    experimental results for $\qty{193}{\giga\electronvolt}$ U+U and $\qty{200}{\giga\electronvolt}$ Au+Au collisions at STAR~\cite{starcollaboration2024imaging}.}
\end{figure}

Finally, we turn to Fig.~\ref{fig:ratio}, which shows the ratio of $v_2\{2\}^2$ between U+U at $\qty{193}{\giga\electronvolt}$ and Au+Au at $\qty{200}{\giga\electronvolt}$, i.e.

\begin{gather}
    \textrm{r}_{\rm Au,U}\left( v_2\{2\}^2 \right) = \frac{v_2\{2\}^2_{\rm U+U}}{v_2\{2\}^2_{\rm Au+Au}}
\end{gather}

We have opted to include only one such ratio, as Fig.~\ref{fig:chargedv22} is sufficiently 
descriptive to enable us to select the most appropriate parametrizations for this calculation, 
namely New $^{238}$U and Def $^{197}$Au (as parametrized in Tab.~\ref{tab:params}), as those 
are obviously closest to the experimental data. The figure shows that our model overestimates
the ratio throughout the considered centrality range. While the ratio is fairly close to the experimental data,
our model's slight underestimation of the Au elliptic flow, discussed previously,
 leads to an overestimation of the ratio.
It is however important to reiterate that our model does very well at reproducing the 
two-particle cumulant of elliptic flow for both 
systems, especially considering that only a single calibration was undertaken.
Furthermore, our calculations between $0-1 \%$ and $3-4\%$ fall close to the experimental values, 
which happen to be the most sensitive points to initial
geometry. Given the excellent agreement between our $v_2\{2\}$ calculation and experimental data,
this result points to 2 possible avenues for improvements. Firstly, it seems like a small increase in 
$\beta_{2, \textrm{Au}}^i$ would help the ratio fall within the 
experimental data over the entire range. Secondly, an individually calibrated run for 
Au may help, although the effects of such a calibration on elliptic flow would be 
small compared to changes in its quadrupole moment.

\subsection{Predictions}

We now move on to predictions of our model for multi-particle correlations and 
transverse-momentum-flow correlations for $\sqrt{s_{NN}} = \qty{193}{\giga\electronvolt}$ 
U+U collisions.

\subsubsection{Differential flow}

\begin{figure}[b]
    \includegraphics[width=\linewidth]{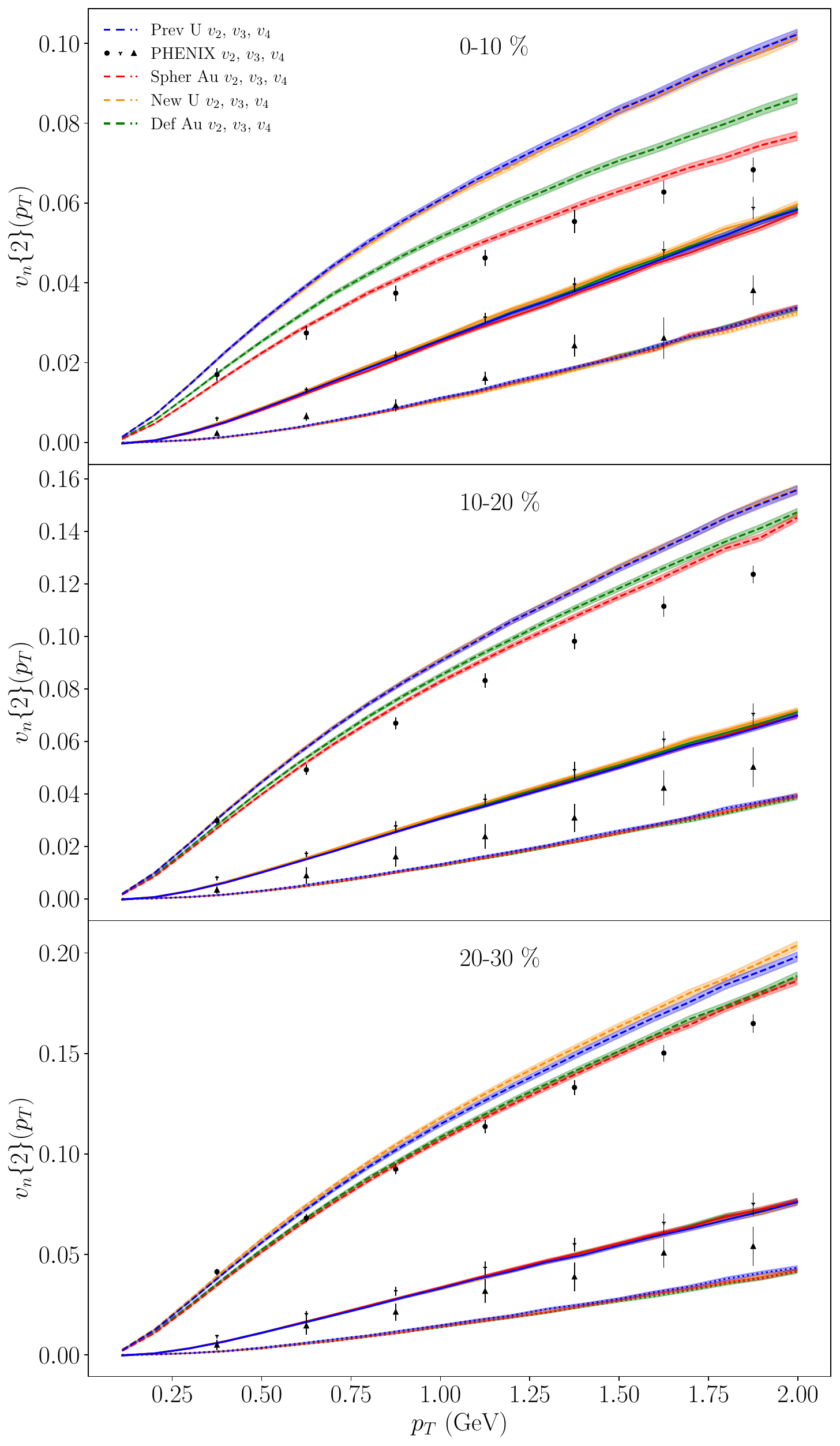}
    \caption{\label{fig:diffcharged} Model predictions of charged hadron differential anisotropic flow 
    coefficients $v_2\{2\}$, $v_3\{2\}$ and $v_4\{2\}$ as functions of transverse 
    momentum $p_T$ for various centrality classes for U+U, 
    compared to experimental results from PHENIX~\cite{Adare_2011} and model calculations for $\qty{200}{\giga\electronvolt}$ Au+Au collisions.}
\end{figure}

In this section, our predictions for U+U collisions will be contextualized using Au+Au experimental 
results and simulation data.

\begin{figure*}[t!]
    \includegraphics[width=\linewidth]{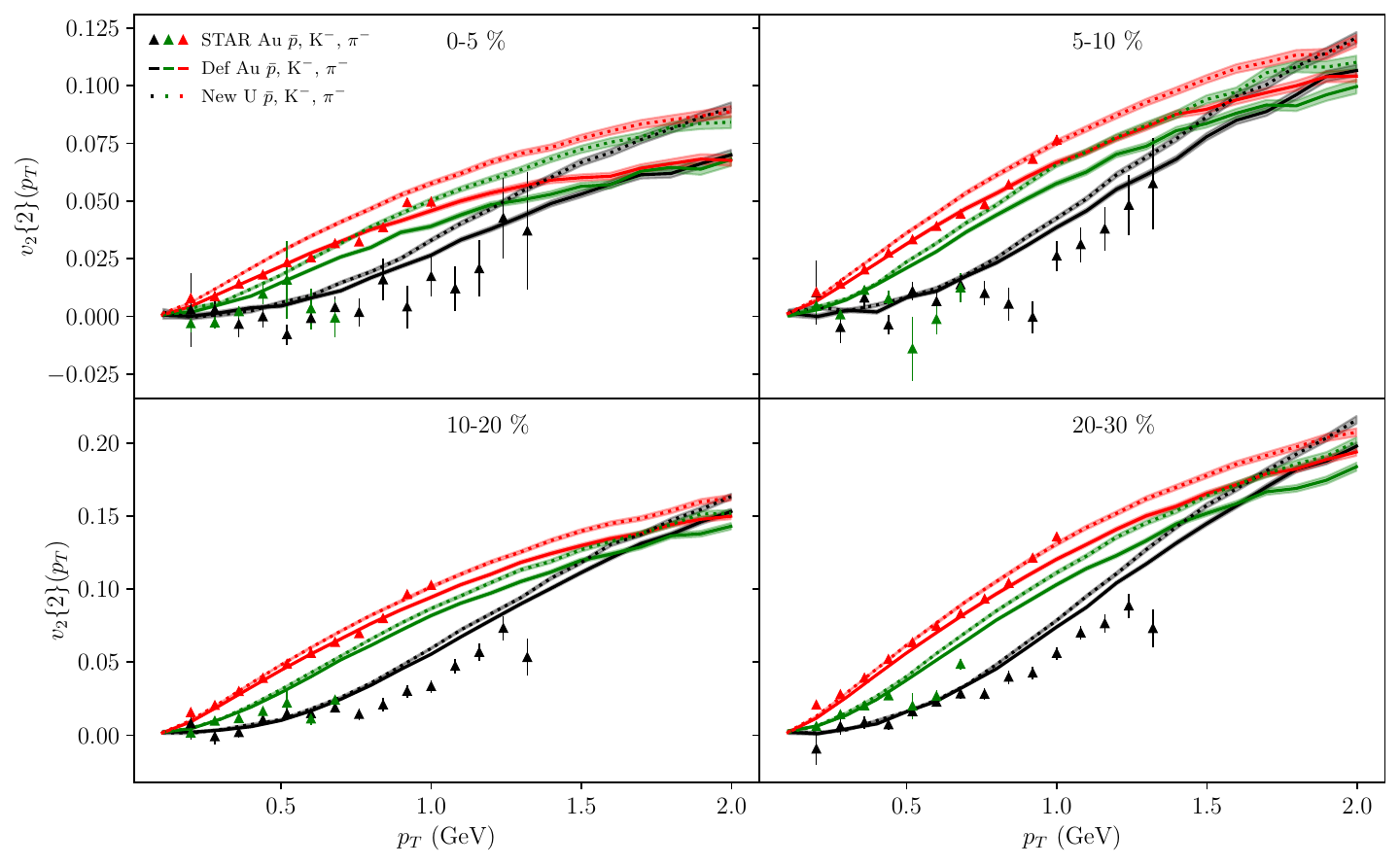}
    \caption{\label{fig:diffidentified} Model prediction of identified particle differential elliptic flow 
    coefficients $v_2\{2\}$ as a function of transverse momentum $p_T$ for various 
    centrality classes for U+U at $\qty{193}{\giga\electronvolt}$, compared to experimental
    results at STAR~\cite{Adams_2005} and model calculations for $\qty{200}{\giga\electronvolt}$ Au+Au collisions.} 
\end{figure*}

Fig.~\ref{fig:diffcharged} introduces our predictions for the differential $v_n$ for U+U at
$\sqrt{s_{NN}} = 193\,\hbox{GeV}$ compared to experimental results and our simulations for Au+Au at 
$\sqrt{s_{NN}} = 200\,\hbox{GeV}$.
We find that our model's calculations of $v_3\{2\}(p_T)$ for U+U and Au+Au
is similar to the Au+Au event-plane $v_3$ data from PHENIX~\cite{Adare_2011}\footnote{The PHENIX data here 
does not use the scalar-product method.}.
This is expected because $v_3$ depends mainly on local fluctuations which are similar in
the two systems, and are unaffected by initial configurations.

Differential elliptic flow ($v_2$) for U+U is larger across all centrality classes than $v_2$ for Au+Au, 
independent of configurations. The differences do, however, get smaller as we move towards
more peripheral regions. For the $0-10 \%$ centrality class,
this is consistent with the deformation effects discussed in section \ref{subsub:ani}: we
expect the elliptic flow to be enhanced in this region of the centrality spectrum because
of the elliptic shape of $^{238}$U nuclei. The two Au configurations overestimate differential
$v_2$ for $p_T \geq \qty{0.5}{\giga\electronvolt}$. While it may seem like cause for concern, it is important to 
remember that most produced particles have transverse momenta less than $\qty{1}{\giga\electronvolt}$, meaning that
higher $p_T$ contributions to the integrated elliptic flow are relatively small. Nevertheless, this high-$p_T$
discrepancy is worth looking into, as it may point towards inadequacies within our hydrodynamics modelling.

In the $10-20 \%$ and $20-30 \%$ classes, the
differences are smaller and are themselves consistent with elliptic flow being noticeably
larger for U+U collisions compared to Au+Au throughout the collision spectrum, as
evidenced in Fig.~\ref{fig:chargedv22}. The two configurations from both systems end up with 
overlapping $v_2$ lines, which further confirms that specific nuclear structure does not play an important
role beyond central collisions. As for $v_4$, both collision systems underestimate this observable
in peripheral ($10-30\%$) collisions. In central ($0-10\%$) collisions, both collision systems overlap with themselves
and the experimental data. Therefore, our model predicts that $v_4$ for U+U collisions at $\qty{193}{\giga\electronvolt}$
should be similar (if not equal) to that of Au+Au collisions at $\qty{200}{\giga\electronvolt}$.

\begin{figure*}[t!]
    \includegraphics[width=\linewidth]{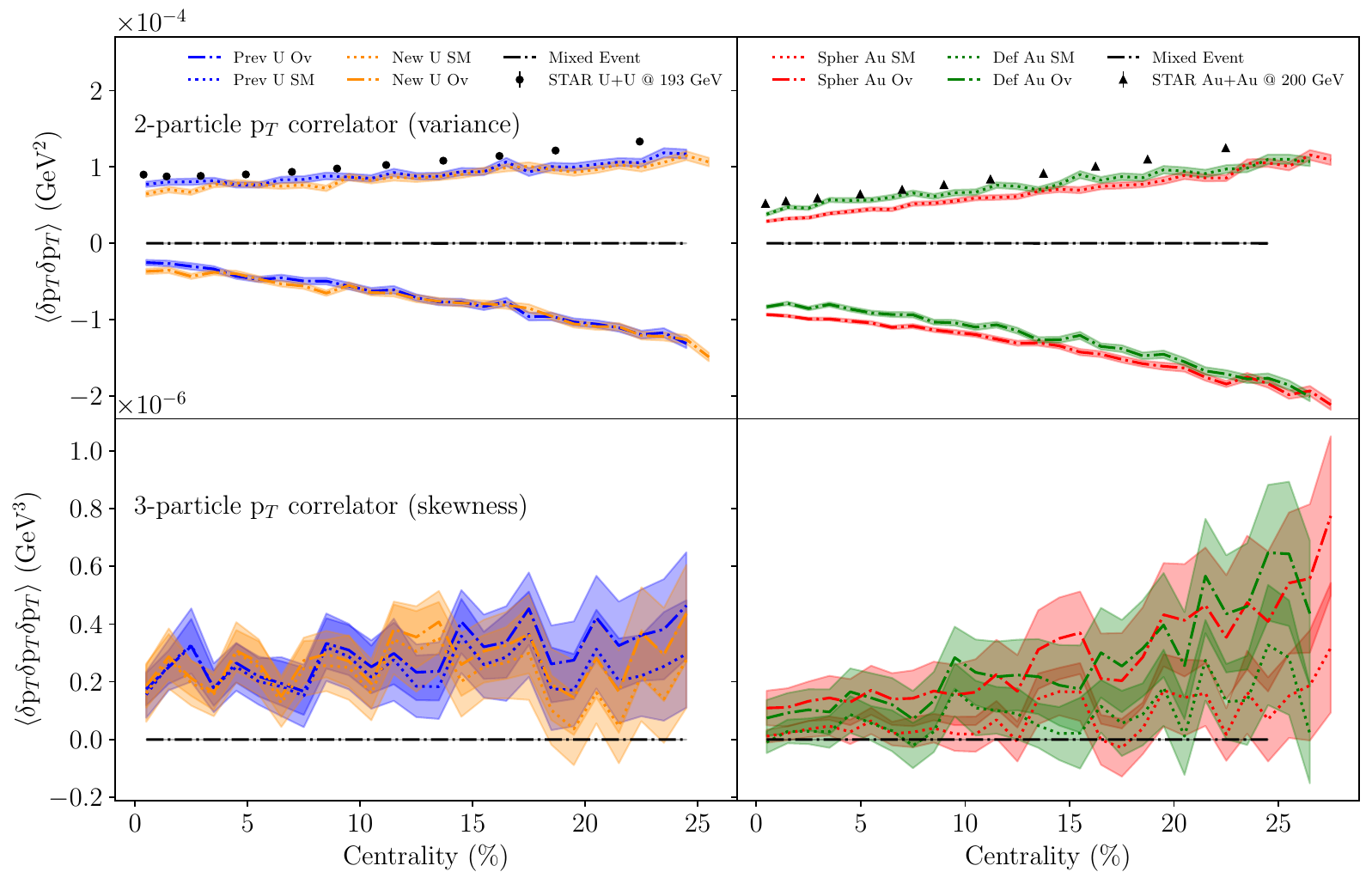}
    \caption{~(\textbf{top}) 2- and (\textbf{bottom}) 3-particle momentum correlators 
    as functions of centrality, with 
    $\qty{0.2}{\giga\electronvolt} \leq p_T \leq \qty{3.0}{\giga\electronvolt}$\label{fig:varskew}, for (\textbf{left}) U+U at $\qty{193}{\giga\electronvolt}$ and (\textbf{right}) Au+Au at $\qty{200}{\giga\electronvolt}$, compared to experimental results at STAR~\cite{starcollaboration2024imaging}. Here, \textit{SM} stands for SMASH sub-event average while \textit{Ov} stands for oversampled average.} 
\end{figure*}

Finally, Fig.~\ref{fig:diffidentified} shows differential elliptic flow for identified 
particles from our model and from Au+Au collisions at STAR
\cite{Adams_2005}. This figure only includes two configurations (one for each system) for clarity,
given that general conclusions given the ordering of the elliptic flow of the different configurations
can already be extracted from Fig.~\ref{fig:diffcharged}. Our Au calculations reproduce the experimental data
well across all four centrality ranges. Our model, as seen previously,
 predicts a larger differential elliptic flow for U than for Au.
In the ultra-central regions ($0-5 \%$ \& $5-10 \%$), the effect is
clear and crosses hadronic lines. However, in the more peripheral regions ($10-20 \%$ \&
$20-30 \%$), this difference becomes much smaller, and varies considerably from one hadron
to another; our model's predictions for anti-protons is similar to both Au
experimental data and our model's calculation. Higher-$p_T$ differential flow seems to converge consistently across
all identified particles.

The final two results subsections will contain predictions of our model compared to experimental data.
This may seem like a contradiction, but our calculations were done and published 
to the arXiv before the publication of Ref.~\cite{starcollaboration2024imaging}, as evidenced by the first version of this manuscript~\cite{fortier2023comparisons}.
As such, the following observables figure as predictions of our model, and no
further calibration was undertaken following the publication of Ref.~\cite{starcollaboration2024imaging}.

\subsubsection{Multi-particle Momentum Correlations}

Multi-particle correlations, when compared to experimental data and combined with
primary observables such as elliptic flow, can help constrain $^{238}$U's, $^{197}$Au's and other
nuclei's deformation parameters and can allow for further analysis using various sets of deformation
parameters~\cite{Jia_2022,starcollaboration2024imaging}.

In the top panels of Fig.~\ref{fig:varskew}, we show the 2-particle $p_T$ correlator 
compared to experimental data from STAR~\cite{starcollaboration2024imaging}. 
In both plots, we have included a mixed event curve to ensure that no underlying $p_T$ correlations 
exist - its position on the plot confirms this. A sizeable difference exists between our 
oversampled event average and SMASH sub-event average curves, which are diametrically opposed
 with respect to our mixed event curve. This indicates that the inclusion (or exclusion) 
 of short-range correlations are a key determinant of the behaviour of this observable across collision systems.
 Our SMASH sub-event average curves match available experimental data rather well, with
 the Prev $^{238}$U and 
 Def $^{197}$Au configurations as best matches. This shows that, contrarily to other primary observables like 
 the elliptic flow where oversampled averaging led to best results, the 2-particle $p_T$ correlator depends on 
 taking the SMASH sub-event average in order to reproduce experimental data. This implies that short-range 
 correlations play a large role in this experimental observable. It is also of interest that the 
 strength of the correlations (or anti-correlations)
 differs between the two collision systems. This difference is mainly driven by deformity, as is evidenced
 by the fact that a clear ordering exists between our configurations, going from least deformed (Spher $^{197}$Au)
 to most deformed (Prev $^{238}$U). Indeed, the experimental U $\fluc{\delta p_T \delta p_T}$ 
 curve increases between $0$ and $2\%$ centrality, while the Au curve does not. Because ultra-central
 collisions of deformed nuclei have a wider variety of geometrical configurations, correlations 
 of dynamical $p_T$ fluctuations increase as well: the number of pairs of particles with $p_T$ considerably
 deviating from the centrality class $\fluc{p_T}$ swells thanks to the distinct groups (body-body and tip-tip)
 of collision geometries present in said centrality class.

The bottom panels of Fig.~\ref{fig:varskew} shows the 3-particle $p_T$ correlator. In
contrast to the upper panel, both curves are extremely similar, with the oversampled
average only slightly larger than the SMASH sub-event averaging values. The oversampled
averaging technique gives out larger correlations, which serve as a further contrast to
the behaviour of the curves in the upper panel. 3-particle correlations are slightly larger, in general,
for U compared to Au, especially in central collisions. Interestingly, the SMASH sub-event averaging
and oversampled averaging curves diverge in peripheral Au collisions; this does not, however, occur
in our U calculations. Once again, the mixed event curve is
included to ensure that no implicit correlations exist for this observable.

\begin{figure}
    \includegraphics[width=\linewidth]{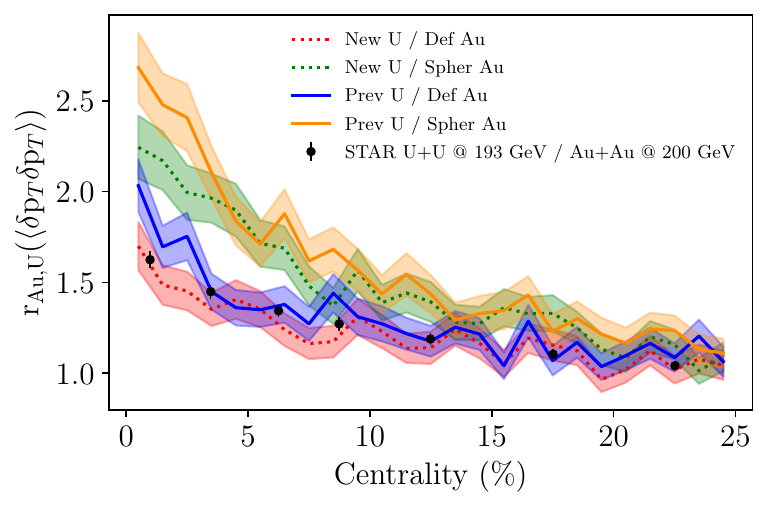}
    \caption{~Ratio of the 2-particle $p_T$ correlator for all U and Au parametrizations, compared to experimental results for U+U at $\qty{193}{\giga\electronvolt}$ and Au+Au at $\qty{200}{\giga\electronvolt}$ collisions at STAR~\cite{starcollaboration2024imaging} \label{fig:var_ratio}} 
\end{figure}

Fig.~\ref{fig:var_ratio} shows the ratio of the 2-particle $p_T$ correlator, defined as 

\begin{gather}
    \textrm{r}_{\rm Au,U}\left(\fluc{\delta p_T \delta p_T}\right) = \frac{\fluc{\delta p_T \delta p_T}_{\rm U}}{\fluc{\delta p_T \delta p_T}_{\rm Au}}
\end{gather}

Here, only the SMASH sub-event averaged correlator values
were used, as they were an obvious match for the available experimental results.
The best ratios are obtained using the Def $^{197}$Au parametrization in the denominator, 
with that using the New $^{238}$U parametrization in the numerator being a slightly better
match for experimental data than Prev $^{238}$U.
This is in contrast to what had been observed in Fig.~\ref{fig:varskew}'s top-left panel, 
where Prev $^{238}$U seemed to have been better at describing the experimental results. This 
shows one of intrinsic cons of using ratios as observables: this new observable is 
really two independent observables wrapped into one, and two underestimates can lead to the ratio being spot-on.

It is however understood that observables measured in one collision system are largely shaped 
by the underlying hydrodynamic properties of the QGP. Taking ratios of observables can, 
in some cases, help ease the existing tension, by giving a composite observable which 
is essentially independent of the hydrodynamic phase~\cite{starcollaboration2024imaging}.

\begin{figure}
    \includegraphics[width=\linewidth]{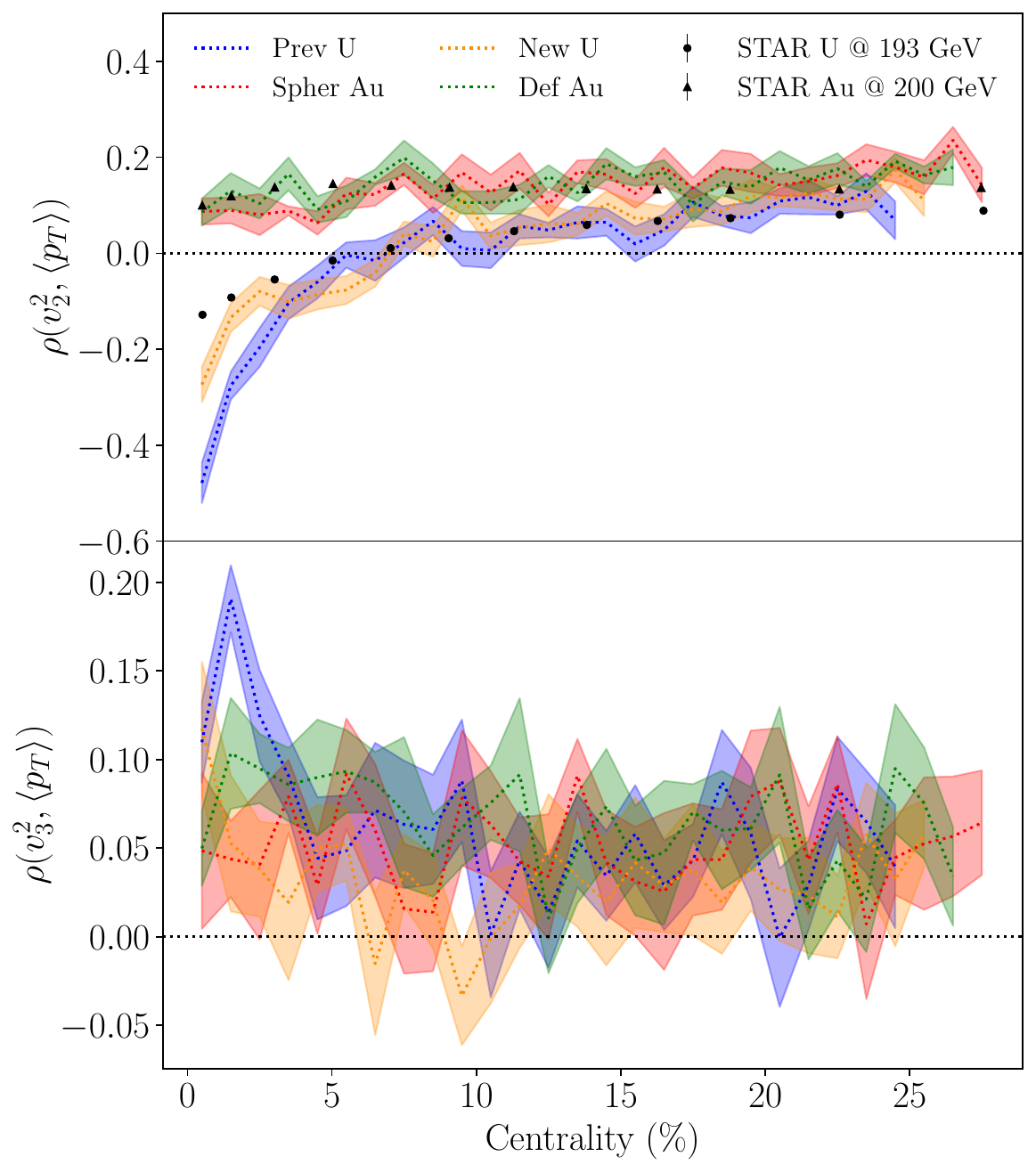}
    \caption{~(\textbf{top}) Elliptic and (\textbf{bottom}) triangular flow and 
    $\fluc{p_T}$ correlations as functions of centrality for U+U at $\qty{193}{\giga\electronvolt}$ and Au+Au at $\qty{200}{\giga\electronvolt}$, compared to experimental results from STAR~\cite{starcollaboration2024imaging}. \label{fig:rho}} 
\end{figure}

\begin{figure}
    \includegraphics[width=\linewidth]{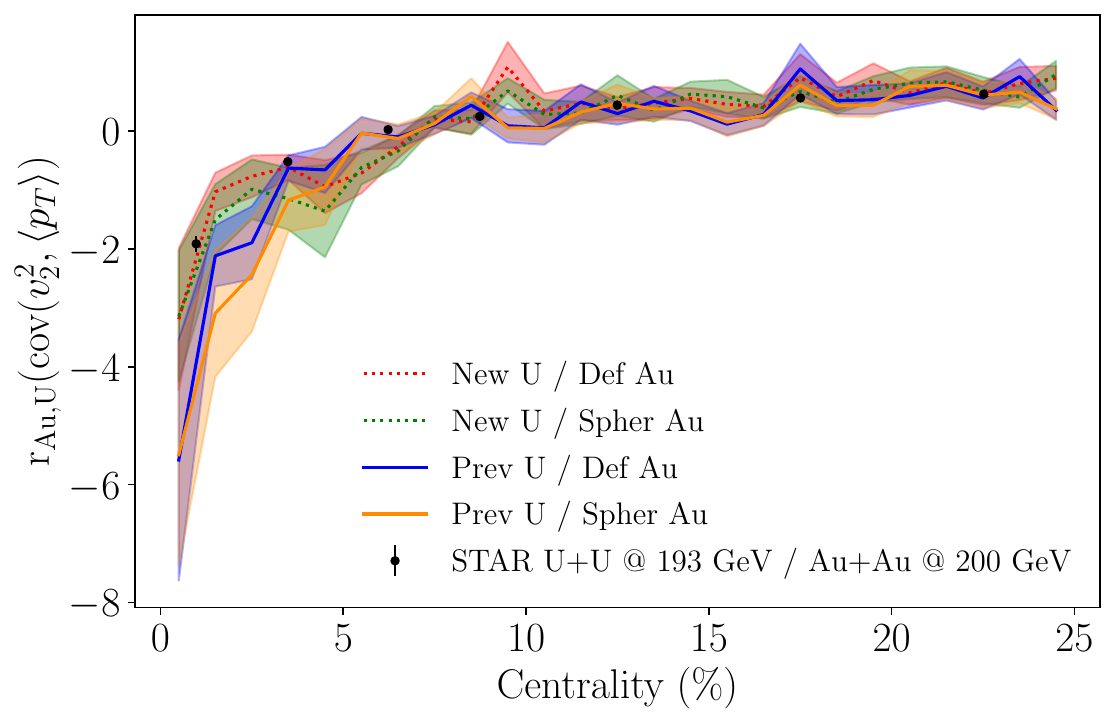}
    \caption{~Ratio of the elliptic-flow-momentum correlations for all U and Au parametrizations, compared to experimental results for U+U at $\qty{193}{\giga\electronvolt}$ and Au+Au at $\qty{200}{\giga\electronvolt}$ collisions at STAR~\cite{starcollaboration2024imaging} \label{fig:rho_ratio}} 
\end{figure}

\subsubsection{Transverse-Momentum-Flow Correlation}

We end our results section with plots showing predictions for the correlations between 
integrated anisotropic flow and mean transverse momentum. This observable has garnered 
interest as a tell-tale sign of deformation~\cite{Giacalone_2021}.

Fig.~\ref{fig:rho} only includes oversampled average calculations, as using the SMASH sub-event averaging
method simply leads us to carry over the overestimation of $v_2\{2\}$ into the correlator.
The mixed event curve could not be included in this plot as it is 
undefined across our centrality interval. Indeed, referring to Eq.\eqref{eq:correlator}, 
the correlator $\rho(v^2_n,\fluc{p_T})$ requires a division by $\fluc{\delta p_T \delta p_T}$
 which is evenly $0$ across our centrality interval, as evidenced by Fig.~\ref{fig:varskew}. 
 Triangular-flow-momentum (bottom panel) correlations seem to be dominated by fluctuations, 
 with the only real clear trend being that the correlator remains positive throughout the 
 centrality range and across both collision systems and nuclear parametrizations. No experimental data was available for this observable. 
 Elliptic-flow-momentum correlations, on the other hand, show clear differences when moving from
 one collision system to the other. Both collision systems achieve excellent
 agreement with available experimental data. Indeed, Au correlations are positive throughout the centrality
 range; they show a slight, consistent increase as we move from central to peripheral collisions. 
 The small quadrupole deformation of Def $^{197}$Au has no tangible effect on the observable.
 U correlations, on the other hand, behave considerably differently. While the effects of changing
 the deformation parameters between runs are as subtle as for Au, a clear cross-over from 
 correlation to anti-correlation occurs at around the $7\%$ centrality mark in both configurations, as well as in the experimental data.

In central collisions of (near-)spherical nuclei~\cite{Schenke_transverse,Aad_2019}, we expect and observe a
dip in the correlator due to the correlation between the inverse of the transverse overlap
area (larger mean $p_T$) and initial state eccentricity $\epsilon_2$ (which gets smaller
in more central collisions). However, no anti-correlation is observed. Our model predicted
that, for U+U collisions, an anti-correlation should be observed in central collisions; the experimental data shows that this prediction turned out true. We can make sense of
this prediction and experimental result by using the correlation between inverse overlap area and eccentricity.
Indeed, tip-tip collisions will generate relatively small overlap areas that are high in
energy density (leading to high $\fluc{p_T}$) while generating almost no eccentricity and
therefore, elliptic flow. Therefore, when contrasted with other events (such as body-body
events) in a given central centrality class, the correlator finds that $\fluc{p_T}$ and
$v_2\{2\}$ are anti-correlated, as events in these classes having lower $\fluc{p_T}$ and
higher eccentricities contrast other events in the same centrality class having larger
$\fluc{p_T}$ and smaller eccentricities.

Finally, as with $v_2\{2\}$ and $\fluc{\delta p_T \delta p_T}$, Fig.~\ref{fig:rho_ratio} shows the ratio of 
covariances of $v_2\{2\}$ and $p_T$ of our two collision systems, i.e.

\begin{gather}
    \textrm{r}_{\rm Au,U}(\operatorname{cov}(v_n\{2\}^2, \fluc{p_T})) = 
    \frac{\operatorname{cov}(v_n\{2\}^2, \fluc{p_T})_{\rm U}}{\operatorname{cov}(v_n\{2\}^2, \fluc{p_T})_{\rm Au}}
\end{gather}

where $\operatorname{cov}(v_n\{2\}^2, \fluc{p_T})$ is defined in Eq.~\eqref{eq:covar}. 
The covariances were kept instead of the full correlation in order to properly
reflect the experimental data found in~\cite{starcollaboration2024imaging}.

All ratios perform well in more peripheral ($> 10\%$ centrality) collisions.
In central collisions, ratios including the Prev $^{238}$U configurations fall noticeably
lower than the experimental results. While both ratios including the New $^{238}$U parametrization perform well, 
the experimental data shows a marked preference for the New $^{238}$U / Def $^{197}$Au ratio. 
Therefore, while Fig.~\ref{fig:rho} could not truly separate between the two Au parametrizations, 
taking the ratio of the covariances has provided additional evidence that a 
deformed Au parametrization is more appropriate (and, therefore, that Au is 
indeed a deformed nucleus).

\section{Summary and Conclusion\label{sec:conc}}

We have presented a detailed synthesis of our multi-phase model consisting of IP-Glasma 
initial state, MUSIC viscous relativistic hydrodynamics, and iS3D + SMASH particle sampling 
and transport, and have shown its results for a wide variety of observables for U+U 
collisions at $\sqrt{s_{NN}} = \qty{193}{\giga\electronvolt}$ and Au+Au at 
$\sqrt{s_{NN}} = \qty{200}{\giga\electronvolt}$.  The purpose of this work is 
to describe the published data and to present timely predictions for observables specifically 
relevant to deformed nuclei. 

Our model shows good agreement across all available experimental data 
for charged hadron multiplicity, identified particle yields and integrated anisotropic flow. 
The underestimation of the $v_3\{2\}$ by about 15\,\% could also be
and opportunity to further study the effects of sub-nucleonic fluctuations.

All-in-all, however, our model performed very well against a limited set
of experimental data and reproducing Au+Au experimental 
data it had not been expressedly calibrated to reproduce. Particularly impressive
was matching new experimental data ($\rho(v_2^2,\fluc{p_T})$,$\fluc{\delta p_T \delta p_T}$) that our model had initially output as predictions.

Our model's physics-based approach allowed us to make predictions regarding flow and bulk
observables and interpret them phenomenologically. Furthermore, we were able to use it to
test different averaging techniques and determine their effects on observables. In the
case of 2-particle momentum correlations, these two methods led to widely different
values. This clear difference presents itself as a golden opportunity to determine which
types of fluctuations and correlations (either local or global) dominate the $p_T$
spectrum in U+U collisions at $\sqrt{s_{NN}} = \qty{193}{\giga\electronvolt}$. This
prediction could also be of use in constraining $^{238}$U deformation parameters, as
momentum correlators are sensitive to nuclear deformation~\cite{Jia_2022}. Another
compelling prediction of our model was that of a definite cross-over towards
anti-correlation of $v_2\{2\}$ and $\fluc{p_T}$ in central collisions, at about the $7 \%$
mark. This is in stark contrast to reported correlations of spherically symmetric nuclei
\cite{Aad_2019}, our calculations for Au+Au at $\sqrt{s_{NN}} = \qty{193}{\giga\electronvolt}$,
as well as results obtained in other works using our model
\cite{heffernan2023bayesian}. If such an anti-correlation is detected in future
experimental work, it would validate our phenomenological reasoning as well as our model.
Finally, our predictions regarding the differential flow for charged hadrons and
identified particles were consistent across centrality classes when compared to data for
Au+Au collisions at RHIC. Indeed, the observables acted as expected given available
integrated flow observables.
We emphasize that even though the experimental data are included for the correlation
observables, our calculations were done before the publication of
Ref.~\cite{starcollaboration2024imaging} as the initial version of this manuscript showed~\cite{fortier2023comparisons}.

Our work here sets the stage for further calculations regarding deformed nuclei, be it
$^{238}$U specifically or others, such as $^{129}$Xe~\cite{Acharya_2019}. Indeed, while
our current work seems to suggest a good understanding of the structure of $^{238}$U and $^{197}$Au, many
details remain to be explored.
As our models become better at describing heavy-ion
collisions, we must prove increasingly diligent regarding the nuclear parametrizations we
use. While our best-performing $^{238}$U and $^{197}$Au parametrizations were satisfactory, 
more details are certainly required to represent physical reality adequately; we think here 
of different and more recent nuclear parametrization paradigms, such as Nuclear Density 
Functional Theory (NDFT)~\cite{doi:10.1080/23746149.2020.1740061}, which provide deeper and more physical
constraints on nucleon positions and correlations. In the end, we would want to use
sensitive observables available experimentally to constrain nuclear deformation,
allowing us to further perfect our state-of-the-art simulations and provide a deeper
understanding of strongly interacting matter at various stages. \\

\section*{Acknowledgments}

We would like to acknowledge the support of the entirety of our research group at McGill
University. We also acknowledge insightful conversations with R.~Modarresi Yazdi, 
M.~Heffernan, S.~McDonald, S.~Shi, B.~Schenke C.~Shen, J.~Jia and C.~Zhang. 
This work was funded by the Natural Sciences and Engineering Research Council
of Canada (NSERC) [SAPIN-2018-00024 ; SAPIN-2020-00048]. Computations were made on the
B\'{e}luga supercomputer system from McGill University, managed by Calcul Qu\'{e}bec
(\url{calculquebec.ca}) and Digital Research Alliance of Canada (\url{alliancecan.ca}).
The operation of this supercomputer is funded by the Canada Foundation for Innovation
(CFI), Minist\`{e}re de l'\'{E}conomie, des Sciences et de l'Innovation du Qu\'{e}bec
(MESI) and le Fonds de recherche du Qu\'{e}bec - Nature et technologies (FRQ-NT).

\bibliography{bibliography}
\end{document}